\documentclass[runningheads]{llncs}
\usepackage{nameref}
\usepackage{multirow}
\usepackage{graphicx}
\usepackage{amsmath}
\usepackage[symbol]{footmisc}
\usepackage[
colorlinks=true,
urlcolor=blue,
linkcolor=red
]{hyperref}
\usepackage{caption,geometry}
\usepackage{float}
\usepackage{longtable}
\begin{document}
\title{Skull-stripping induces shortcut learning in MRI-based Alzheimer's disease classification}
\titlerunning{Skull-stripping induces shortcut learning in MRI-based Alzheimer's disease classification}
%
\author{Christian Tinauer\inst{1}\orcidID{0000-0003-4355-3898} \and
Maximilian Sackl\inst{1}\orcidID{0000-0002-1927-7953} \and
Rudolf Stollberger\inst{2,3}\orcidID{0000-0002-4969-3878} \and
Reinhold Schmidt\inst{1}\orcidID{0000-0002-6406-7584} \and
Stefan Ropele\inst{1,3}\orcidID{0000-0002-5559-768X} \and
Christian Langkammer\inst{1,3}\orcidID{0000-0002-7097-9707},\\
for the Alzheimer's Disease Neuroimaging Initiative\inst{*}}

\footnotetext[1]{Data used in preparation of this article were obtained from the Alzheimer's Disease Neuroimaging Initiative (ADNI) database (\url{https://adni.loni.usc.edu}). As such, the investigators within the ADNI contributed to the design and implementation of ADNI and/or provided data but did not participate in analysis or writing of this report. A complete listing of ADNI investigators can be found at: \url{https://adni.loni.usc.edu/wp-content/uploads/how_to_apply/ADNI_Acknowledgement_List.pdf}}

\authorrunning{Tinauer et al.}
%
\institute{Department of Neurology, Medical University of Graz, 8036 Graz, Austria
\email{\{christian.tinauer,christian.langkammer\}@medunigraz.at}\\
\url{https://neuroimaging.at} \and
Institute of Biomedical Imaging, Graz University of Technology, 8010 Graz, Austria \and
BioTechMed-Graz, 8010 Graz, Austria}

\maketitle              
\begin{abstract}
\hfill\break
\textbf{\emph{Objectives.}} High classification accuracy of Alzheimer's disease (AD) from structural MRI has been achieved using deep neural networks, yet the specific image features contributing to these decisions remain unclear. In this study, the contributions of T1-weighted (T1w) gray-white matter texture, volumetric information, and preprocessing -particularly skull-stripping- were systematically assessed.
\\~\\
\textbf{\emph{Methods.}} A dataset of 990 matched T1w MRIs from AD patients and cognitively normal controls from the ADNI database were used. Preprocessing was varied through skull-stripping and intensity binarization to isolate texture and shape contributions. A 3D convolutional neural network was trained on each configuration, and classification performance was compared using exact McNemar tests with discrete Bonferroni-Holm correction. Feature relevance was analyzed using Layer-wise Relevance Propagation, image similarity metrics, and spectral clustering of relevance maps.
\\~\\
\textbf{\emph{Results.}} Despite substantial differences in image content, classification accuracy, sensitivity, and specificity remained stable across preprocessing conditions. Models trained on binarized images preserved performance, indicating minimal reliance on gray-white matter texture. Instead, volumetric features -particularly brain contours introduced through skull-stripping- were consistently used by the models.
\\~\\
\textbf{\emph{Conclusions.}} This behavior reflects a shortcut learning phenomenon, where preprocessing artifacts act as potentially unintended cues. The resulting Clever Hans effect emphasizes the critical importance of interpretability tools to reveal hidden biases and to ensure robust and trustworthy deep learning in medical imaging.
\\~\\
\textbf{\emph{Relevance Statement.}} We investigated the mechanisms underlying deep learning-based disease classification using a widely utilized Alzheimer's disease dataset and our findings reveal a reliance on features induced through skull-stripping, highlighting the need for careful preprocessing to ensure clinically relevant and interpretable models.
\\~\\
\textbf{\emph{Key Points.}} 
\begin{itemize}
    \item Demonstrates that shortcut learning is induced by skull-stripping applied to T1-weighted MRIs.
    \item Uses McNemar tests, explainable deep learning and spectral clustering to estimate the bias.
    \item Highlights the importance of understanding the dataset, image preprocessing and deep learning model, for interpretation and validation.
\end{itemize}

\end{abstract}
\section{Introduction}
Alzheimer's disease (AD) is the most common form of dementia, accounting for 60-70\% of cases \cite{scheltens_alzheimers_2021} 
and with over 55 million people worldwide living with some form of dementia, it poses a substantial burden 
on healthcare systems, caregivers, and families \cite{noauthor_2024_2024}. However, in vivo diagnosis remains challenging due to 
the overlap of clinical symptoms with other conditions, resulting in relatively low diagnostic accuracy 
(71-87\% sensitivity and 44-71\% specificity) \cite{beach_accuracy_2012}.
\\~\\
In addition to clinical and neuropsychological assessments, medical imaging is employed to improve diagnostic accuracy. 
Positron emission tomography (PET) imaging with amyloid and tau protein ligands, combined with magnetic resonance imaging (MRI), 
has become a valuable tool in AD diagnosis \cite{dubois_clinical_2021}. Yet, AD is characterized by a prolonged prodromal and asymptomatic inflammatory phase, 
during which PET imaging is unsuitable for predicting disease onset in healthy populations. Since pathological changes in AD begin decades 
before clinical symptoms appear, MRI holds promise for identifying early biomarkers in a broad population. Currently, brain volumetry \cite{sluimer_whole-brain_2008}, 
particularly hippocampal atrophy \cite{henneman_hippocampal_2009}, is widely used as an imaging marker for differential diagnosis and in interventional studies.
\\~\\
In recent years, convolutional neural networks (CNNs) have emerged as the state-of-the-art method for AD classification using 
structural T1-weighted (T1w) MRI scans \cite{wen_convolutional_2020}. These networks learn image features directly during the training process, eliminating 
the need for manual feature selection. Despite their advantages, CNN models and the features they extract are often difficult for humans 
to interpret, earning them the reputation of being "black boxes" \cite{davatzikos_machine_2019}.
\\~\\
To address this issue, interpretability methods like heatmapping have been introduced \cite{simonyan_deep_2014}. One notable technique is Layer-wise Relevance Propagation (LRP) \cite{bach_pixel-wise_2015}, which highlights input features driving model decisions. The importance of such tools is illustrated by cases like \cite{lapuschkin_unmasking_2019}, where a model identified horses based on watermarks rather than the animals - an instance of the Clever Hans effect \cite{bottani_evaluation_2023,wallis_clever_2022}. This term, drawn from a horse once thought to perform arithmetic but later shown to respond to subtle cues \cite{pfungst_clever_1911}, exemplifies shortcut learning: the exploitation of spurious correlations over meaningful patterns \cite{geirhos_shortcut_2020}.
\\~\\
In AD classification, interpretability research has shown that preprocessing steps can shape both performance and learned features \cite{tinauer_interpretable_2022}. We hypothesize that skull-stripping, a common step, may introduce unintended cues and lead models to overlook more relevant AD-specific markers, such as structural atrophy and gray-white matter contrast changes \cite{canu_mapping_2011}.
\\~\\
In this study, we combined deep learning and heatmapping explainability techniques to evaluate the performance and learned features of CNNs trained on different input configurations. Specifically, we trained identical CNN architectures on full T1w images from the Alzheimer's Disease Neuroimaging Initiative (ADNI) dataset, their skull-stripped counterparts, and three differently binarized versions of these two preprocessing approaches, creating a total of eight model configurations. We first analyzed performance metrics for significant differences, assessed structural similarities between LRP-based heatmaps, and investigated the spatial distribution of heatmap relevance to examine whether preprocessing introduced unintended cues \cite{vasquez-venegas_detecting_2024}.

\section{Materials and Methods}
\subsection{Imaging Data}
Using MR image metadata from the ADNI database (\url{https://adni.loni.usc.edu}), we created a subset of images with clinically available and consistent properties. The final search criteria were: Phase = ADNI 2, Acquisition Plane = SAGITTAL, Field Strength = 3.0 Tesla, Pixel Spacing XY = [1.0 mm, 1.1 mm], Slice Thickness = 1.2 mm, and Weighting = T1. These criteria resulted in 1,042 images for the AD group and 2,227 images for the normal control (NC) group. This ensured that patients and controls were scanned using a consistent MRI protocol at 3 Tesla across multiple scanner vendors and sites. Supplementary Table~\ref{tab:binADNI_supplemental_1_pivot_dim_site_manufacturer} and Supplementary Table~\ref{tab:binADNI_supplemental_1_pivot_dim_manufacturer_protocol} in \nameref{sec:sm_1} provide an overview of the distribution of images across sites, vendors, imaging protocols, and research groups.

\subsection{Research Groups}
We retrospectively selected 990 MR images from 159 patients with AD and 990 MR images from 201 propensity-logit-matched NCs that were propensity-logit-matched using age and sex as covariates from the preselected image data subset \cite{kline_psmpy_2022}. Supplementary Figure~\ref{fig:effect_sizes} in \nameref{sec:sm_1} details effect sizes before and after matching. Table~\ref{tab:demographics} shows the demographics of the selection. Data were split into training, validation, and test sets (ratio 70:15:15), ensuring that all scans from a given individual were included in the same set. To maintain consistent class distribution, the final sets were created by combining data from both cohorts.

\begin{table}
\caption{Summary of subject demographics at baseline for ADNI. Note. Values are presented as mean ± SD [range]. M: male, F: female, MMSE: mini-mental state examination, CDR: global clinical dementia rating, APOE: Apolipoprotein E status, Education in years, n/a: no value available.}
\label{tab:demographics}
\resizebox{\textwidth}{!}{%
\begin{tabular}{|l|l|l|l|l|l|l|l|l|}
\hline
   & \textbf{Subjects} & \textbf{Images} & \textbf{Age}                                                        & \textbf{Gender}                                       & \textbf{MMSE}                                                                 & \textbf{CDR}                                                                    & \textbf{APOE}                                                                                                               & \textbf{Education}                                                           \\ \hline \hline
NC & 201               & 990             & \begin{tabular}[c]{@{}l@{}}75.1±7.1\\ {[}56.3, 95.8{]}\end{tabular} & \begin{tabular}[c]{@{}l@{}}102 M/\\ 99 F\end{tabular} & \begin{tabular}[c]{@{}l@{}}28.9±1.2\\ {[}24.0, 30.0{]}\\ n/a: 16\end{tabular} & \begin{tabular}[c]{@{}l@{}}0.0: 171;\\ 0.5: 13;\\ n/a: 17\end{tabular}          & \begin{tabular}[c]{@{}l@{}}$\epsilon$2/$\epsilon$2: 1;\\ $\epsilon$2/$\epsilon$3: 20;\\ $\epsilon$2/$\epsilon$4: 1;\\ $\epsilon$3/$\epsilon$3: 112;\\ $\epsilon$3/$\epsilon$4: 49;\\ $\epsilon$4/$\epsilon$4: 5;\\ n/a: 13\end{tabular} & \begin{tabular}[c]{@{}l@{}}16.7±2.5\\ {[}12.0, 20.0{]}\\ n/a: 0\end{tabular} \\ \hline
AD & 159               & 990             & \begin{tabular}[c]{@{}l@{}}75.3±7.9\\ {[}55.7, 91.5{]}\end{tabular} & \begin{tabular}[c]{@{}l@{}}91 M/\\ 68 F\end{tabular}  & \begin{tabular}[c]{@{}l@{}}22.0±3.8\\ {[}4.0, 30.0{]}\\ n/a: 48\end{tabular}  & \begin{tabular}[c]{@{}l@{}}0.5: 35;\\ 1.0: 67;\\ 2.0: 9;\\ n/a: 48\end{tabular} & \begin{tabular}[c]{@{}l@{}}$\epsilon$2/$\epsilon$2: 1;\\ $\epsilon$2/$\epsilon$3: 5;\\ $\epsilon$2/$\epsilon$4: 2;\\ $\epsilon$3/$\epsilon$3: 41;\\ $\epsilon$3/$\epsilon$4: 68;\\ $\epsilon$4/$\epsilon$4: 29;\\ n/a: 13\end{tabular}  & \begin{tabular}[c]{@{}l@{}}15.7±2.7\\ {[}9.0, 20.0{]}\\ n/a: 0\end{tabular}  \\ \hline
\end{tabular}
}
\end{table}

\subsection{Preprocessing}
\begin{figure}
    \includegraphics[width=\textwidth]{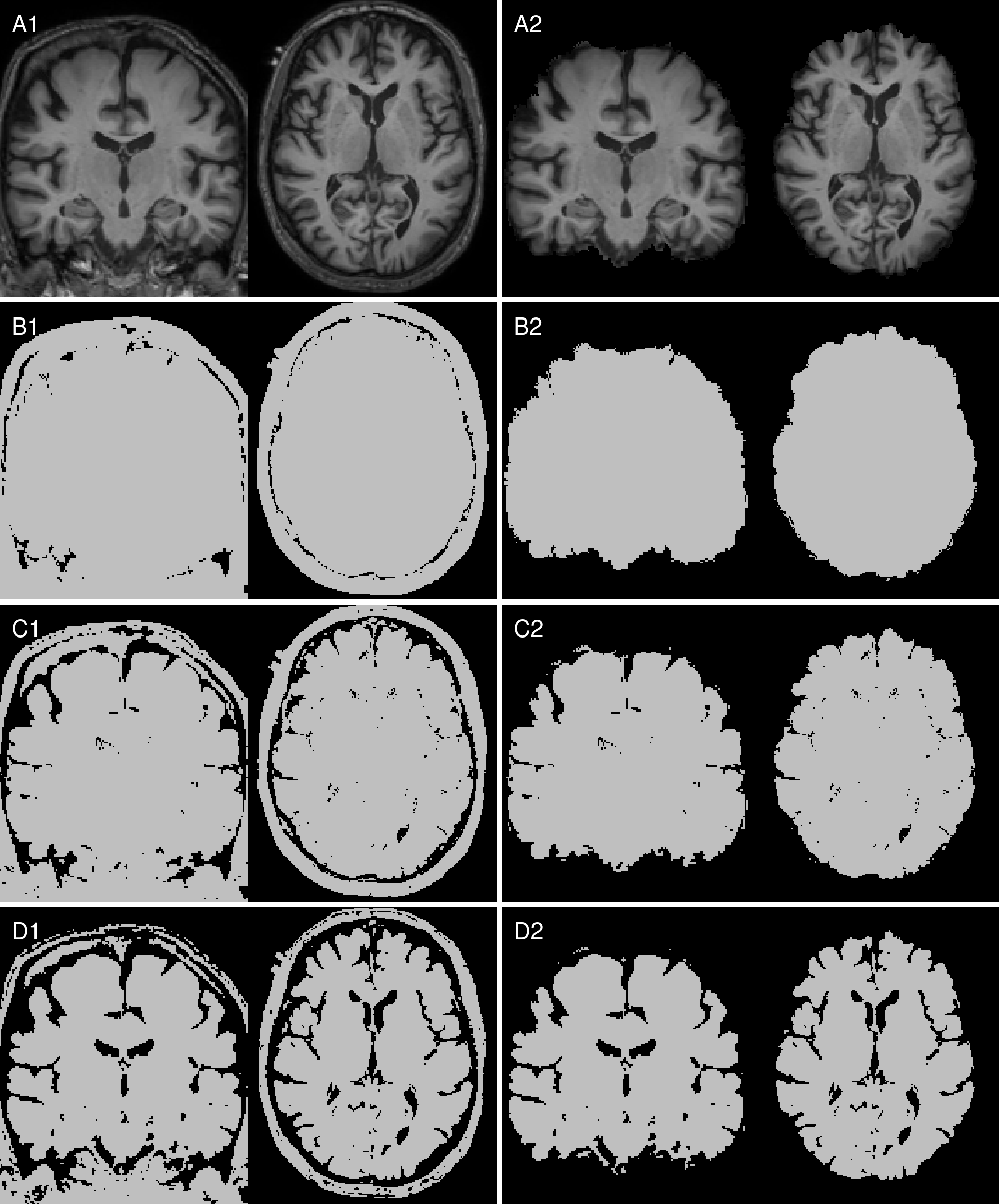}
    \caption{Input image setups: Left column with (A1) aligned T1w MRI, identical binarized T1w images with the manually 
    selected threshold levels (B1) 13.75\%, (C1) 27.50\% (C1) and (D1) 41.25\%, and in the right column the corresponding 
    skull-stripped versions (A2, B2, C2, D2).}
    \label{fig:example}
\end{figure}

Raw T1w images were reoriented to standard space using FSL-REORIENT2STD \cite{jenkinson_fsl_2012}, cropped to a $160\times240\times256$ matrix size, bias field corrected using N4 \cite{tustison_n4itk_2010}, and non-linearly registered to the MNI152 template via FSL-FNIRT \cite{jenkinson_fsl_2012}. Intensity values were normalized to the white matter peak of the brain tissue histogram (196 bins). The outputs of this preprocessing pipeline are referred to as “aligned” images. Individual brain masks were generated in native image space using SIENAX from FSL \cite{smith_accurate_2002} and warped to the aligned images to create the “skull-stripped” images. Binary masks preserving shape information were derived using manually selected thresholds of 13.75\%, 27.5\%, and 41.25\% of the white matter peak of the brain tissue histogram and the aligned images. These binary masks were also combined with skull-stripped preprocessing, resulting in eight total setups, as illustrated in Figure~\ref{fig:example}. Supplementary Figure~\ref{fig:group_divergence} in \nameref{sec:sm_1} illustrates that thresholds were selected to preserve meaningful atrophy patterns by comparing residual voxels with individual brain masks.

\subsection{Standard classification network}
\begin{figure}
    \includegraphics[width=\textwidth]{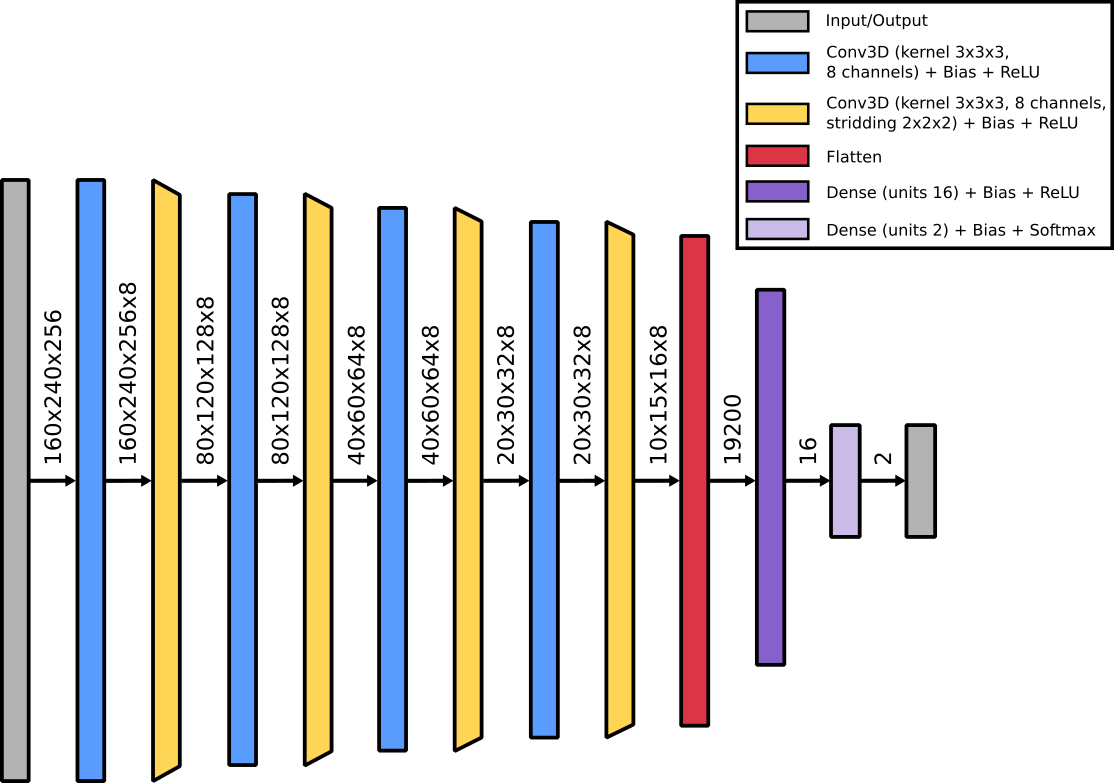}
    \caption{Structure of the 3D classifier network. Dimensionalities between layers are the tensor sizes.}
    \label{fig:classifier}
\end{figure}

We utilized a conventional 3D subject-level classifier network as described in \cite{wen_convolutional_2020}. However, because the number of trainable parameters (42 million) relative to the dataset size (1980 images) is high, the network is prone to overfitting. To address this, we reduced the number and size of the convolutional and fully connected layers until the network no longer overfit the training data and the validation accuracy ceased to improve. See \nameref{sec:sm_2} for loss and performance curves. Batch normalization layers did not influence the network's performance and were therefore omitted. Additionally, we replaced max pooling layers with convolutional layers using striding, as tested in \cite{springenberg_striving_2015}. This modification improves the interpretability of the network \cite{montavon_methods_2018}. Dropout was not applied in the network. To further enhance interpretability, all biases in the classifier were constrained to be non-positive, which helped sparsify the network activations \cite{montavon_methods_2018}.
\\~\\
The final 3D classifier network, as shown in Figure~\ref{fig:classifier}, consists of a single convolutional layer (kernel size: $3\times3\times3$, 8 channels) combined with a down-convolutional layer (kernel size: $3\times3\times3$, 8 channels, striding: 2) as its primary building block. The network stacks four of these main building blocks, followed by two fully connected layers (with 16 and 2 units, respectively), resulting in a total of 0.3 million trainable parameters. Each layer is followed by a Rectified Linear Unit activation function, except for the output layer, which employs a Softmax activation.

\subsection{Training}
Models were trained on aligned images, skull-stripped images, and their binarized counterparts. 
Training was performed using the Adam optimizer \cite{kingma_adam_2015} for 30 epochs with a batch size of 20 using three independent 
network weights initializations \cite{bouthillier_accounting_2021} minimizing the binary cross entropy loss. 
Each data sampling was trained with all three initializations.

\subsection{Heatmapping and relevance-weighted heatmap presentation}
Heatmaps were created using the LRP method with $\alpha$=1.0 and $\beta$=0.0, as described in \cite{bach_pixel-wise_2015}. 
Each voxel is attributed a relevance score (R). To analyze relevance heatmaps, we qualitatively examined 
individual maps and calculated mean heatmaps for each configuration. For each mean heatmap, we generated a 
histogram of relevance values. Starting from the bin with the highest relevance, bin contents were iteratively 
summed until 40\% of the total relevance was included. The lower bound of the final bin was used as the lower 
threshold for windowing the mean heatmap. All heatmaps presented in this study are overlaid on the MNI152 1 mm 
template and windowed to display the top 40\% of relevance.

\subsection{Random sampling}
For each data sampling (10 samplings) and each network weight initialization (3 initializations), 
we retrained the network, resulting in 30 training sessions for each input image configuration \cite{bradshaw_guide_2023}. 
Non-converging training sessions were excluded from further analysis. From the remaining sessions, 
the best-performing run based on validation accuracy was selected to create the mean heatmaps.

\subsection{Statistical analysis}
We calculated performance metrics, including accuracy, sensitivity, specificity, and the area under the receiver operating characteristic curve (AUC), 
for all model configurations. Results are reported as mean values, standard deviations, and 95\% confidence intervals.
The model trained on skull-stripped T1w images (see A2 in Figure~\ref{fig:example}) was designated as the reference model for statistical comparisons. 
Exact McNemar tests were performed to compare accuracy, sensitivity, and specificity between the reference model and each 
alternative model for each session run (combination of data sampling and weight initialization), provided both runs were available \cite{dietterich_approximate_1998}. 
To account for multiple comparisons, we applied the discrete Bonferroni-Holm correction, a statistical method suitable for adjusting comparisons involving overlapping data splits \cite{westfall_multiple_2010}.

\subsection{Heatmap similarity analysis}
To evaluate heatmap similarities, we compared the best-performing run of the reference model with the corresponding runs (same data sampling and weight initialization) of each alternative model using structural image similarity measures. Before comparison heatmaps were normalized to min-max. Voxel-wise similarity was assessed with the root mean square error (RMSE), while global trends and overall similarity were evaluated using the Pearson correlation coefficient. Both global and localized patterns were analyzed with the mean structural similarity index measure (MSSIM) and Earth Mover's Distance (EMD). Additionally, binarized heatmaps highlighting the top 40\% and top 10\% relevance values were compared using the Intersection over Union (IoU) metric.

\section{Results}
The following sections present quantitative performance results across setups, followed by an analysis of CNN-extracted features using LRP and image similarity metrics.

\subsection{Model performances}
Table~\ref{tab:performance} provides the results of the accuracy, sensitivity, specificity, and AUC for all configurations in the random 
sampling setup. Performance metrics were evaluated across non-excluded training sessions for each model configuration.
\\~\\
Using an alpha level of .05 and discrete Bonferroni-Holm correction for multiple exact McNemar tests, comparisons between the reference model 
(A2, skull-stripped, no binarization) and alternative configurations -including binarized skull-stripped images at thresholds of 13.75\% 
(B2, 8 significant differences), 27.50\% (C2, 1 significant difference), and 41.25\% (D2, 4 significant differences) as well as aligned 
images binarized at 41.25\% (D1, 7 significant differences)- revealed little to no evidence for significant differences in accuracy, 
sensitivity, and specificity. 120 of 600 total comparisons remained significant after correction for multiple testing. 
See Supplementary Table~\ref{tab:statistical_comparison} in \nameref{sec:sm_3} for significant differences in model comparisons.
\\~\\
Aligned images with binarization thresholds 13.75\% and 27.50\% (B1, C1) performed comparably to aligned T1w images without binarization (A1). Similarly, skull-stripped images with binarization (B2, C2, D2) and aligned images binarized at 41.25\% (D1) exhibited comparable performance to skull-stripped images without binarization (A2) while outperforming other aligned image configurations.
\\~\\
Additionally, model performances were tested on our local, non-public datasets. See \nameref{sec:sm_4} for results.

\begin{table}
    \caption{Summary of performance metrics of all configurations. Note. AUC = area under receiver operating characteristics curve. Values between [ and ] show the 95\% confidence interval.}
    \label{tab:performance}
    \resizebox{\textwidth}{!}{%
    \begin{tabular}{|l|l|l|l|l|l|l|}
    \hline
    \textbf{Input images}               & \textbf{Id} & \textbf{Binarizer} & \textbf{Accuracy}                                                                      & \textbf{Sensitivity}                                                                   & \textbf{Specificity}                                                                   & \textbf{AUC}                                                                   \\ \hline \hline
    \multirow{7}{*}{Aligned T1w}        & A1          & None               & \begin{tabular}[c]{@{}l@{}}71.12±5.01\%\\ {[}61.34\%, 82.52\%{]}\end{tabular}          & \begin{tabular}[c]{@{}l@{}}67.47±9.90\%\\ {[}51.66\%, 85.94\%{]}\end{tabular}          & \begin{tabular}[c]{@{}l@{}}74.76±7.07\%\\ {[}62.03\%, 85.45\%{]}\end{tabular}          & \begin{tabular}[c]{@{}l@{}}0.71±0.05\\ {[}0.62, 0.83{]}\end{tabular}           \\ \cline{2-7} 
                                        & B1          & 13.75\%            & \begin{tabular}[c]{@{}l@{}}62.51±5.45\%\\ {[}53.39\%, 72.09\%{]}\end{tabular}          & \begin{tabular}[c]{@{}l@{}}62.26±9.48\%\\ {[}47.70\%, 84.93\%{]}\end{tabular}          & \begin{tabular}[c]{@{}l@{}}62.79±8.35\%\\ {[}46.56\%, 74.42\%{]}\end{tabular}          & \begin{tabular}[c]{@{}l@{}}0.63±0.054\\ {[}0.53, 0.72{]}\end{tabular}          \\ \cline{2-7} 
                                        & C1          & 27.50\%            & \begin{tabular}[c]{@{}l@{}}72.74±5.49\%\\ {[}61.41\%, 82.11\%{]}\end{tabular}          & \begin{tabular}[c]{@{}l@{}}71.37±9.10\%\\ {[}56.86\%, 88.87\%{]}\end{tabular}          & \begin{tabular}[c]{@{}l@{}}74.15±9.94\%\\ {[}51.67\%, 88.37\%{]}\end{tabular}          & \begin{tabular}[c]{@{}l@{}}0.73±0.055\\ {[}0.61, 0.82{]}\end{tabular}          \\ \cline{2-7} 
                                        & D1          & 41.25\%            & \begin{tabular}[c]{@{}l@{}}77.95±4.57\%\\ {[}70.90\%, 86.34\%{]}\end{tabular}          & \begin{tabular}[c]{@{}l@{}}76.74±9.41\%\\ {[}60.10\%, 94.56\%{]}\end{tabular}          & \begin{tabular}[c]{@{}l@{}}79.15±6.80\%\\ {[}64.54\%, 89.56\%{]}\end{tabular}          & \begin{tabular}[c]{@{}l@{}}0.78±0.045\\ {[}0.71, 0.86\end{tabular}             \\ \hline
    \multirow{7}{*}{Skull-stripped T1w} & A2          & None               & \textbf{\begin{tabular}[c]{@{}l@{}}81.63±3.77\%\\ {[}74.36\%, 88.01\%{]}\end{tabular}} & \textbf{\begin{tabular}[c]{@{}l@{}}81.22±6.94\%\\ {[}69.59\%, 93.15\%{]}\end{tabular}} & \begin{tabular}[c]{@{}l@{}}82.11±7.92\%\\ {[}65.50\%, 93.60\%{]}\end{tabular}          & \textbf{\begin{tabular}[c]{@{}l@{}}0.82±0.037\\ {[}0.74, 0.88{]}\end{tabular}} \\ \cline{2-7} 
                                        & B2          & 13.75\%            & \begin{tabular}[c]{@{}l@{}}78.12±4.63\%\\ {[}70.79\%, 85.79\%{]}\end{tabular}          & \begin{tabular}[c]{@{}l@{}}76.83±7.03\%\\ {[}62.65\%, 87.05\%{]}\end{tabular}          & \begin{tabular}[c]{@{}l@{}}79.40±6.76\%\\ {[}65.53\%, 89.91\%{]}\end{tabular}          & \begin{tabular}[c]{@{}l@{}}0.78±0.046\\ {[}0.71, 0.86{]}\end{tabular}          \\ \cline{2-7} 
                                        & C2          & 27.50\%            & \begin{tabular}[c]{@{}l@{}}79.57±3.92\%\\ {[}73.46\%, 86.45\%{]}\end{tabular}          & \begin{tabular}[c]{@{}l@{}}78.32±7.74\%\\ {[}66.79\%, 93.87\%{]}\end{tabular}          & \begin{tabular}[c]{@{}l@{}}80.92±7.71\%\\ {[}67.53\%, 91.95\%{]}\end{tabular}          & \begin{tabular}[c]{@{}l@{}}0.80±0.039\\ {[}0.74, 0.86{]}\end{tabular}          \\ \cline{2-7} 
                                        & D2          & 41.25\%            & \begin{tabular}[c]{@{}l@{}}81.56±4.63\%\\ {[}72.31\%, 88.67\%{]}\end{tabular}          & \begin{tabular}[c]{@{}l@{}}79.69±9.42\%\\ {[}62.59\%, 96.48\%{]}\end{tabular}          & \textbf{\begin{tabular}[c]{@{}l@{}}83.50±6.77\%\\ {[}72.48\%, 96.29\%{]}\end{tabular}} & \begin{tabular}[c]{@{}l@{}}0.82±0.046\\ {[}0.72, 0.89{]}\end{tabular}          \\ \hline
    \end{tabular}%
    }
\end{table}

\subsection{Feature similarities}
\begin{figure}
    \includegraphics[width=\textwidth]{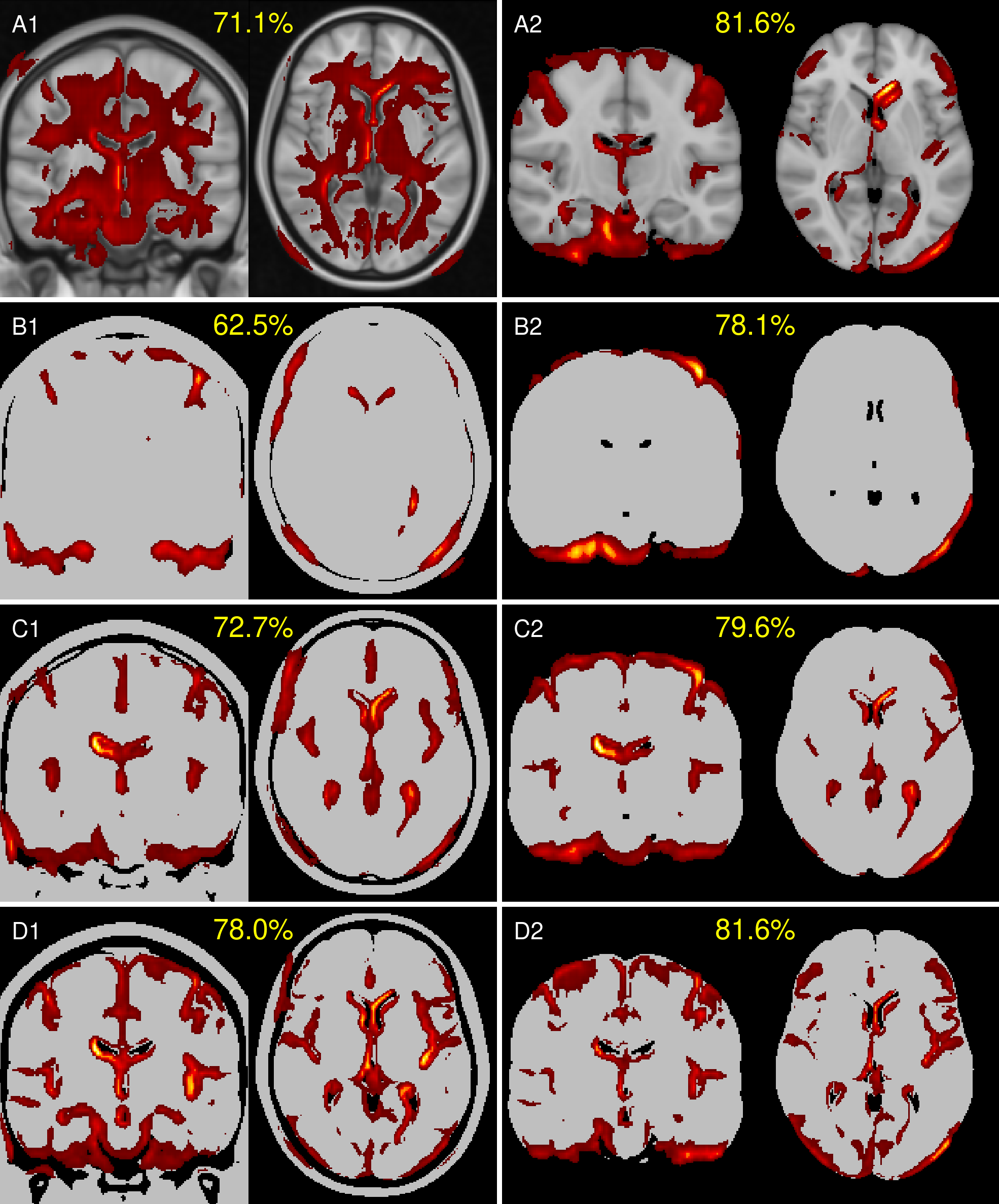}
    \caption{Mean heatmaps from test images: Left column with (A1) aligned T1w MRI, identical 
    binarized T1w image with threshold levels (B1) 13.75\%, (C1) 27.50\% (C1) and (D1) 41.25\%, and right column with corresponding 
    skull-stripped versions (A2, B2, C2, D2). The mean accuracies of the models are shown in yellow.}
    \label{fig:heatmaps}
\end{figure}

Figure~\ref{fig:heatmaps} shows mean heatmaps for classification decisions on test images. Skull-stripping enhances classification accuracy, while mean heatmaps from binarized inputs closely resemble those from non-binarized inputs.
\\~\\
Heatmaps from the skull-stripped model (A2) serve as the reference for structural heatmap comparisons in Table 3. Skull-stripped binarization (B2, C2, D2) shows lower voxel-wise dissimilarity (RMSE), higher global similarity (Pearson Correlation), and improved localized similarity (MSSIM, EMD) compared to the aligned versions (A1, B1, C1, D1). Among the binarized models, skull-stripped-binarization-13.75\% (B2) demonstrates the highest overall structural similarity (RMSE, Pearson Correlation, MSSIM, EMD) with the reference (A2), while skull-stripped-binarization-27.50\% exhibits the strongest regional overlap with the reference. These results emphasize the dominant role of volumetric features over T1w contrast variations. Features similarities and model misclassification analysis were furthermore done using spectral clustering \cite{lapuschkin_unmasking_2019,von_luxburg_tutorial_2007} and t-distributed stochastic neighbor embedding \cite{maaten_visualizing_2008}. See \nameref{sec:sm_5} for an introduction and results.

\begin{table}
    \caption{Summary of heatmap similarity metrics for all configurations compared to the reference model (A2). Note. Values between [ and ] show the 95\% confidence interval. Bin. = Binarizer, RMSE = root mean square error, Corr. = Correlation, MSSIM = mean structural similarity index measure, EMD = Earth Mover's Distance, IoU = Intersection over Union, R = Relevance.}
    \label{tab:similarity}
    \resizebox{\textwidth}{!}{%
    \begin{tabular}{|l|l|l|l|l|l|l|l|l|}
    \hline
    \textbf{Input images}                                                          & \textbf{Id} & \textbf{Bin. \%} & \textbf{RMSE}                                                                 & \textbf{Pearson Corr.}                                                        & \textbf{MSSIM}                                                                & \textbf{EMD}                                                                  & \textbf{\begin{tabular}[c]{@{}l@{}}IoU\\ Top 40\% R\end{tabular}}             & \textbf{\begin{tabular}[c]{@{}l@{}}IoU\\ Top 10\% R\end{tabular}}             \\ \hline \hline
    \multirow{7}{*}{Aligned T1w}                                                   & A1          & None             & \begin{tabular}[c]{@{}l@{}}10.78±3.76\\ {[}5.98, 21.47{]}\end{tabular}        & \begin{tabular}[c]{@{}l@{}}0.13±0.04\\ {[}0.06, 0.21{]}\end{tabular}          & \begin{tabular}[c]{@{}l@{}}0.25±0.08\\ {[}0.12, 0.42{]}\end{tabular}          & \begin{tabular}[c]{@{}l@{}}8.44±3.25\\ {[}3.99, 17.48{]}\end{tabular}         & \begin{tabular}[c]{@{}l@{}}0.06±0.02\\ {[}0.02, 0.11{]}\end{tabular}          & \begin{tabular}[c]{@{}l@{}}0.01±0.01\\ {[}0.00, 0.03{]}\end{tabular}          \\ \cline{2-9} 
                                                                                   & B1          & 13.75            & \begin{tabular}[c]{@{}l@{}}3.85±0.87\\ {[}2.25, 5.48{]}\end{tabular}          & \begin{tabular}[c]{@{}l@{}}0.08±0.03\\ {[}0.03, 0.15{]}\end{tabular}          & \begin{tabular}[c]{@{}l@{}}0.87±0.04\\ {[}0.78, 0.94{]}\end{tabular}          & \begin{tabular}[c]{@{}l@{}}0.50±0.29\\ {[}0.14, 1.29{]}\end{tabular}          & \begin{tabular}[c]{@{}l@{}}0.02±0.01\\ {[}0.01, 0.04{]}\end{tabular}          & \begin{tabular}[c]{@{}l@{}}0.00±0.00\\ {[}0.00, 0.01{]}\end{tabular}          \\ \cline{2-9} 
                                                                                   & C1          & 27.50            & \begin{tabular}[c]{@{}l@{}}4.77±0.96\\ {[}3.07, 6.64{]}\end{tabular}          & \begin{tabular}[c]{@{}l@{}}0.18±0.05\\ {[}0.08, 0.26{]}\end{tabular}          & \begin{tabular}[c]{@{}l@{}}0.83±0.04\\ {[}0.74, 0.89{]}\end{tabular}          & \begin{tabular}[c]{@{}l@{}}0.71±0.45\\ {[}0.11, 1.71{]}\end{tabular}          & \begin{tabular}[c]{@{}l@{}}0.07±0.02\\ {[}0.03, 0.10{]}\end{tabular}          & \begin{tabular}[c]{@{}l@{}}0.01±0.01\\ {[}0.00, 0.02{]}\end{tabular}          \\ \cline{2-9} 
                                                                                   & D1          & 41.25            & \begin{tabular}[c]{@{}l@{}}6.17±1.15\\ {[}3.98, 8.34{]}\end{tabular}          & \begin{tabular}[c]{@{}l@{}}0.23±0.05\\ {[}0.13, 0.30{]}\end{tabular}          & \begin{tabular}[c]{@{}l@{}}0.72±0.06\\ {[}0.62, 0.83{]}\end{tabular}          & \begin{tabular}[c]{@{}l@{}}1.61±0.62\\ {[}0.54, 2.75{]}\end{tabular}          & \begin{tabular}[c]{@{}l@{}}0.10±0.02\\ {[}0.05, 0.14{]}\end{tabular}          & \begin{tabular}[c]{@{}l@{}}0.01±0.01\\ {[}0.00, 0.03{]}\end{tabular}          \\ \hline
    \multirow{7}{*}{\begin{tabular}[c]{@{}l@{}}Skull-\\ stripped T1w\end{tabular}} & A2          & None             & 0.00±0.00                                                                     & 1.00±0.00                                                                     & 1.00±0.00                                                                     & 0.00±0.00                                                                     & 1.00±0.00                                                                     & 1.00±0.00                                                                     \\ \cline{2-9} 
                                                                                   & B2          & 13.75            & \textbf{\begin{tabular}[c]{@{}l@{}}2.79±0.49\\ {[}1.97, 3.79{]}\end{tabular}} & \textbf{\begin{tabular}[c]{@{}l@{}}0.58±0.07\\ {[}0.44, 0.69{]}\end{tabular}} & \textbf{\begin{tabular}[c]{@{}l@{}}0.91±0.03\\ {[}0.82, 0.96{]}\end{tabular}} & \textbf{\begin{tabular}[c]{@{}l@{}}0.39±0.25\\ {[}0.12, 1.05{]}\end{tabular}} & \begin{tabular}[c]{@{}l@{}}0.24±0.06\\ {[}0.13, 0.35{]}\end{tabular}          & \begin{tabular}[c]{@{}l@{}}0.15±0.03\\ {[}0.08, 0.21{]}\end{tabular}          \\ \cline{2-9} 
                                                                                   & C2          & 27.50            & \begin{tabular}[c]{@{}l@{}}3.54±0.61\\ {[}2.45, 4.85{]}\end{tabular}          & \begin{tabular}[c]{@{}l@{}}0.56±0.07\\ {[}0.39, 0.67{]}\end{tabular}          & \begin{tabular}[c]{@{}l@{}}0.86±0.04\\ {[}0.78, 0.92{]}\end{tabular}          & \begin{tabular}[c]{@{}l@{}}0.73±0.44\\ {[}0.13, 1.67{]}\end{tabular}          & \textbf{\begin{tabular}[c]{@{}l@{}}0.31±0.05\\ {[}0.20, 0.37{]}\end{tabular}} & \textbf{\begin{tabular}[c]{@{}l@{}}0.17±0.03\\ {[}0.10, 0.24{]}\end{tabular}} \\ \cline{2-9} 
                                                                                   & D2          & 41.25            & \begin{tabular}[c]{@{}l@{}}3.72±0.59\\ {[}2.69, 4.96{]}\end{tabular}          & \begin{tabular}[c]{@{}l@{}}0.51±0.08\\ {[}0.33, 0.65{]}\end{tabular}          & \begin{tabular}[c]{@{}l@{}}0.80±0.05\\ {[}0.68, 0.89{]}\end{tabular}          & \begin{tabular}[c]{@{}l@{}}1.01±0.50\\ {[}0.23, 2.12{]}\end{tabular}          & \begin{tabular}[c]{@{}l@{}}0.26±0.03\\ {[}0.18, 0.31{]}\end{tabular}          & \begin{tabular}[c]{@{}l@{}}0.16±0.04\\ {[}0.09, 0.22{]}\end{tabular}          \\ \hline
    \end{tabular}%
    }
\end{table}

\section{Discussion}
Previous studies using T1w MRI for AD classification have reported strong CNN performance but often failed to clarify which image features -such as volumetric patterns, signal textures, or preprocessing artifacts- drive model decisions \cite{tinauer_interpretable_2022,bohle_layer-wise_2019,serra_grey_2010}. This lack of interpretability raises concerns about spurious correlations and shortcut learning. We address this gap through a systematic analysis using 1,980 T1w MR images from a widely used AD dataset. Applying explainable deep learning, McNemar-based model comparisons, and spectral clustering of LRP heatmaps, we disentangled the contributions of intensity and anatomical information. Texture removal via image binarization further isolated structural cues. Our analysis reveals a bias: CNNs rely heavily on volume-based features rather than biologically specific microstructural changes. These findings highlight the need for rigorous validation of AI tools and a deeper understanding of how dataset properties, preprocessing, and model behavior interact - an integrative perspective still lacking in the field.
\\~\\
To enable robust and unbiased evaluation, we curated a dataset with high inter-class image similarity and balanced class proportions \cite{wen_convolutional_2020}. We then applied one-to-one matching by sex and age to reduce covariate-driven influences, aiming to minimize confounders and ensure models focused on AD-related structural and contrast changes \cite{canu_mapping_2011,leonardsen_constructing_2024}.
\\~\\
All models used non-linear registration to align MRIs to MNI152 space, accounting for individual anatomical variation. However, this does not fully correct for atrophy, which varies regionally in AD, especially in the hippocampus and cortical gray matter \cite{tinauer_interpretable_2022}.
\\~\\
Skull-stripping, our reference preprocessing step, is widely used in AD classification and supported by heatmap-based studies showing improved accuracy in AD \cite{bohle_layer-wise_2019}, dementia \cite{leonardsen_constructing_2024}, multiple sclerosis \cite{eitel_uncovering_2019}, and brain age prediction \cite{hofmann_towards_2022,hofmann_utility_2025,dinsdale_learning_2021}.
\\~\\
Heatmaps generated in this study corroborate these findings, with relevance predominantly concentrated at the tissue boundary. To further probe the contribution of volumetric features, we removed tissue contrast entirely through binarization, isolating atrophic and structural features for analysis.
\\~\\
We investigated three binarization levels, each aligned with the white matter intensity peak of the image, as depicted in Figure~\ref{fig:example}. These levels, although chosen arbitrarily, retained differing proportions of anatomical structures, capturing distinct aspects of atrophy, including ventricular enlargement, hippocampal shrinking and cerebellum morphology. As CNNs seem to focus on high-contrast regions \cite{eitel_uncovering_2019,mattia_investigating_2024}, these binarization levels allowed us to systematically dissect how different volumetric and structural features influenced model predictions.
\\~\\
Performance metrics, summarized in Table~\ref{tab:performance}, reveal that removing gray-white matter contrast while retaining skull-stripping has little to no effect on model performance. Statistical analyses using exact McNemar tests, adjusted with discrete Bonferroni-Holm correction ($\alpha=.05$), revealed minimal evidence of significant differences in accuracy, sensitivity, and specificity across configurations when compared to the reference model. Specifically, skull-stripped and binarized models at thresholds of 13.75\%, 27.50\%, and 41.25\%, as well as aligned images binarized at 41.25\%, showed fewer than 10\% of comparisons with significant differences. This suggests that volumetric information is sufficient for achieving high classification accuracy in CNN-based AD classification, with minimal contribution from gray-white texture variations.
\\~\\
Given the consistent model performances across configurations, we examined structural image similarities using similarity metrics applied on heatmaps, shown in Table~\ref{tab:similarity}. Surprisingly, the model trained with binarization at 13.75\% (B2) -which retains the most tissue within the brain mask- exhibited the highest similarity to the reference model (A2) across global (RMSE, Pearson correlation) and localized (MSSIM, EMD) similarity metrics. This indicates that the key volumetric and morphological features driving classification are predominantly encoded in the brain mask's volume and shape.
\\~\\
Furthermore, starting at binarization-27.50\% (C1, C2), models began to incorporate additional regions such as the ventricles, with binarization at 41.25\% (D1, D2) capturing also hippocampal features. Notably, the highest overlaps between reference and binarized models were observed at 27.50\% (C2), as indicated by intersection-over-union metrics. This suggests that contrast in ventricular regions in the reference model provides sufficient signal for the model to identify disease-relevant patterns.
\\~\\
Overall is the classification performance of the models driven by high contrast variations, either given by brain structures like the ventricular system and the hippocampi, or by being introduced artificially through image preprocessing. Multi-center studies have repeatedly shown that scanner vendor, acquisition protocol, and preprocessing pipeline systematically alter radiomic and image features, and consequently classifier behavior \cite{kushol_effects_2023,bhagwat_understanding_2021}. Effects can persist even after harmonization if applied improperly \cite{lu_evaluation_2025,marzi_efficacy_2024}.
\\~\\
Drawing parallels to the Clever Hans effect -where unintended cues in the experimental setup were inadvertently learned- we observed a similar phenomenon in deep learning-based AD classification. Preprocessing, particularly skull-stripping in T1w imaging, is crucial for achieving state-of-the-art performance \cite{wen_convolutional_2020,bohle_layer-wise_2019,leonardsen_constructing_2024,dinsdale_learning_2021}, as demonstrated by the inferior results of model A1 (aligned, no binarization) compared to A2 (skull-stripped, no binarization). However, when combined with a CNN, this preprocessing acts as an interviewer effect, steering the model toward artificially introduced yet well-established features like volumetry. This underscores the need for careful control of preprocessing artifacts \cite{tinauer_interpretable_2022} and suggests that quantitative MRI parameter maps \cite{tinauer_explainable_2024,tinauer_identifying_2025,malhi_quantitative_2025} or model regularization \cite{tinauer_interpretable_2022} could offer a more robust alternative by minimizing reliance on such artificial cues.

\subsection{Limitations}
This study has its limitations. First, while the dataset is large, representative, and carefully crafted, it is derived solely from the ADNI database. However, the final models were tested on our local, nonpublic datasets. Second, our analysis primarily examines T1w images, preprocessing strategies, their impact on classification performance, and the features extracted using heatmaps. Third, the CNN architecture was intentionally simplified to control overfitting and was optimized for the reference model (skull-stripped, no binarization) by systematically reducing trainable parameters and complexity, alongside hyperparameter tuning. Although the final architecture achieved performance metrics comparable to existing literature, applying the same setup across experiments aimed to minimize bias but may not eliminate it entirely. It remains to be investigated whether similar effects would arise in more complex 3D architectures. However, we expect that such models would also be susceptible to the bias estimated in this study. Lastly, while heatmaps and similarity metrics were effective for feature interpretation, they may not fully capture the intricate interactions between features learned by the models.

\subsection{Conclusion}
Our findings uncover a shortcut learning effect in deep learning-driven AD classification, demonstrating that models predominantly rely on volumetric features rather than microstructural changes in gray- and white matter. This highlights the critical need to evaluate data selection and preprocessing workflows to distinguish between artifacts and true disease-specific patterns, ensuring clinical relevance.
\\~\\
The implications of this work extend beyond AD classification, urging to adopt robust strategies for disentangling artifacts from meaningful features in deep learning workflows. Model validation pipelines should routinely test model sensitivity to preprocessing variations and data handling choices before clinical deployment. Future studies should incorporate quantitative MRI parameters, such as T1, R2*, or QSM, to provide insights into disease pathology and enhance model interpretability and generalizability. By addressing these challenges, the field can advance toward more reliable and clinically actionable applications of AI in neuroimaging.

\begin{credits}
\subsubsection{Data and Code Availability.}
The code and the image ids (ADNI images) used in this study are available under \url{https://github.com/christiantinauer/binADNI}. The MR images from our local datasets are not publicly available. Formal data sharing requests to the corresponding authors will be considered.

\subsubsection{Ethics Statement.}
Data used in this study were obtained from the Alzheimer’s Disease Neuroimaging Initiative (ADNI) database (\url{https://adni.loni.usc.edu}). The ADNI study was conducted in accordance with the ethical standards of each participating institution’s Institutional Review Board and with the 1964 Helsinki declaration and its later amendments. Written informed consent was obtained from all participants or their authorized representatives at the time of enrollment.
\\~\\
The current study involved only secondary analyses of fully de-identified data provided by ADNI, and no additional ethical approval was required. All analyses were conducted in accordance with the ADNI Data Use Agreement.
\\~\\
Additionally, this study used data acquired in local studies approved by the ethics committee of the Medical University of Graz (IRB00002556) and signed informed consent was obtained from all study participants or their caregivers. The trial protocol for this prospective study was registered at the National Library of Medicine (trial identification number: NCT02752750). All methods were performed in accordance with the relevant guidelines and regulations.

\subsubsection{Funding.}
This study was funded by the Austrian Science Fund (FWF grant numbers: P30134, P35887).

\subsubsection{Declaration of Competing Interest.}
The authors declare no competing interests.

\subsubsection{\ackname}
We thank Lukas Pirpamer for the feedback provided during the creation of this manuscript.
\\~\\
Data collection and sharing for this project was funded by the Alzheimer's Disease Neuroimaging Initiative (ADNI) (National Institutes of Health Grant U01 AG024904) and DOD ADNI (Department of Defense award number W81XWH-12-2-0012). ADNI is funded by the National Institute on Aging, the National Institute of Biomedical Imaging and Bioengineering, and through generous contributions from the following: AbbVie, Alzheimer's Association; Alzheimer's Drug Discovery Foundation; Araclon Biotech; BioClinica, Inc.; Biogen; Bristol-Myers Squibb Company; CereSpir, Inc.; Cogstate; Eisai Inc.; Elan Pharmaceuticals, Inc.; Eli Lilly and Company; EuroImmun; F. Hoffmann-La Roche Ltd and its affiliated company Genentech, Inc.; Fujirebio; GE Healthcare; IXICO Ltd.; Janssen Alzheimer Immunotherapy Research \& Development, LLC.; Johnson \& Johnson Pharmaceutical Research \& Development LLC.; Lumosity; Lundbeck; Merck \& Co., Inc.; Meso Scale Diagnostics, LLC.; NeuroRx Research; Neurotrack Technologies; Novartis Pharmaceuticals Corporation; Pfizer Inc.; Piramal Imaging; Servier; Takeda Pharmaceutical Company; and Transition Therapeutics. The Canadian Institutes of Health Research is providing funds to support ADNI clinical sites in Canada. Private sector contributions are facilitated by the Foundation for the National Institutes of Health (www.fnih.org). The grantee organization is the Northern California Institute for Research and Education, and the study is coordinated by the Alzheimer's Therapeutic Research Institute at the University of Southern California. ADNI data are disseminated by the Laboratory for Neuro Imaging at the University of Southern California.
\\~\\
ChatGPT (model GPT-4o) was used in January, March and July 2025 to assist with grammatics and to streamline parts of the text after initial drafting but we take full responsibility for the content of the manuscript.

\end{credits}

%
%
%
\bibliographystyle{splncs04}
\bibliography{main}

\clearpage
\section*{Supplementary Material 1}
\label{sec:sm_1}

Table~\ref{tab:binADNI_supplemental_1_pivot_dim_site_manufacturer} presents the distribution of preselected images across sites, vendors, and research groups, while Table~\ref{tab:binADNI_supplemental_1_pivot_dim_manufacturer_protocol} details their distribution across vendors, imaging protocols, and research groups.

\newgeometry{scale=1}
\thispagestyle{empty}
{%
  \captionof{table}{Distribution of preselected images by site, vendor and research group.}
  \label{tab:binADNI_supplemental_1_pivot_dim_site_manufacturer}
  \includegraphics[page=1,scale=.85]{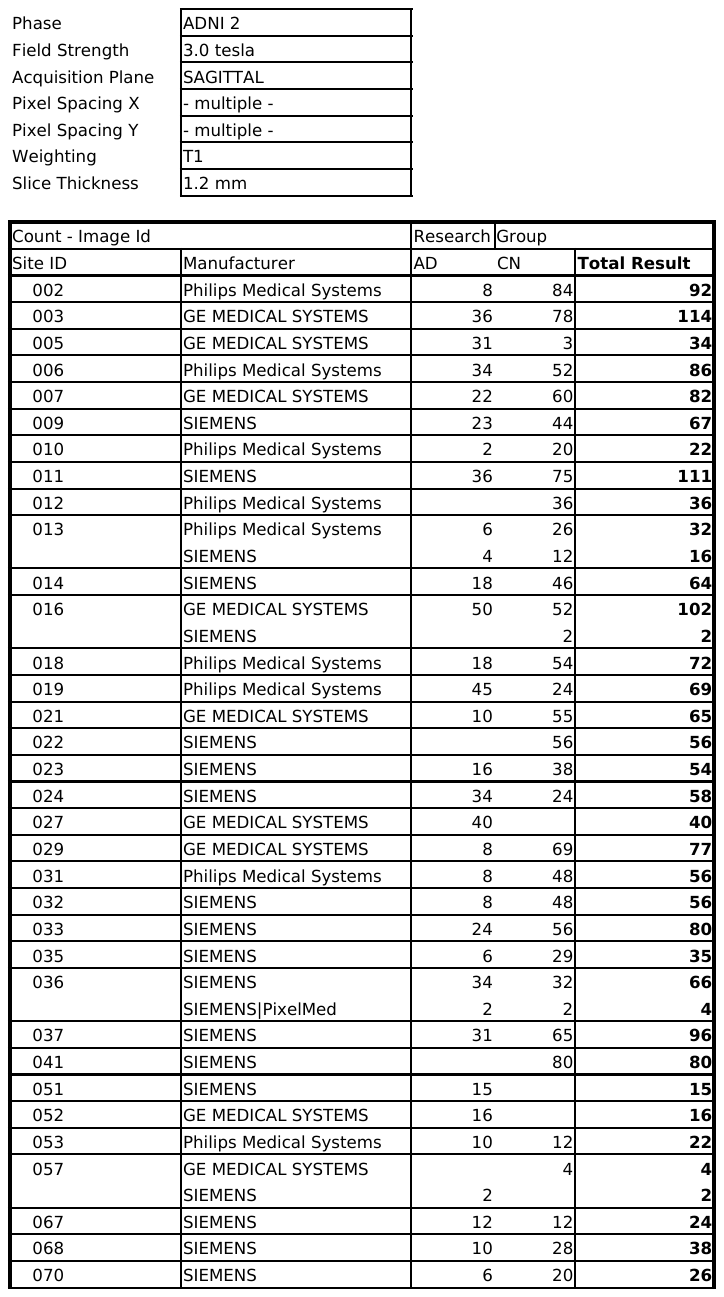}
  \par
}
\restoregeometry

\newgeometry{scale=1}
\thispagestyle{empty}
{%
  \includegraphics[page=2,scale=.85]{pivots/ADNI2_Pivot_dim_Site_Manufacturer.pdf}
  \par
}
\restoregeometry

\newgeometry{scale=1}
\thispagestyle{empty}
{%
  \captionof{table}{Distribution of preselected images by vendor, imaging protocol and research group.}
  \label{tab:binADNI_supplemental_1_pivot_dim_manufacturer_protocol}
  \includegraphics[page=1,scale=.85]{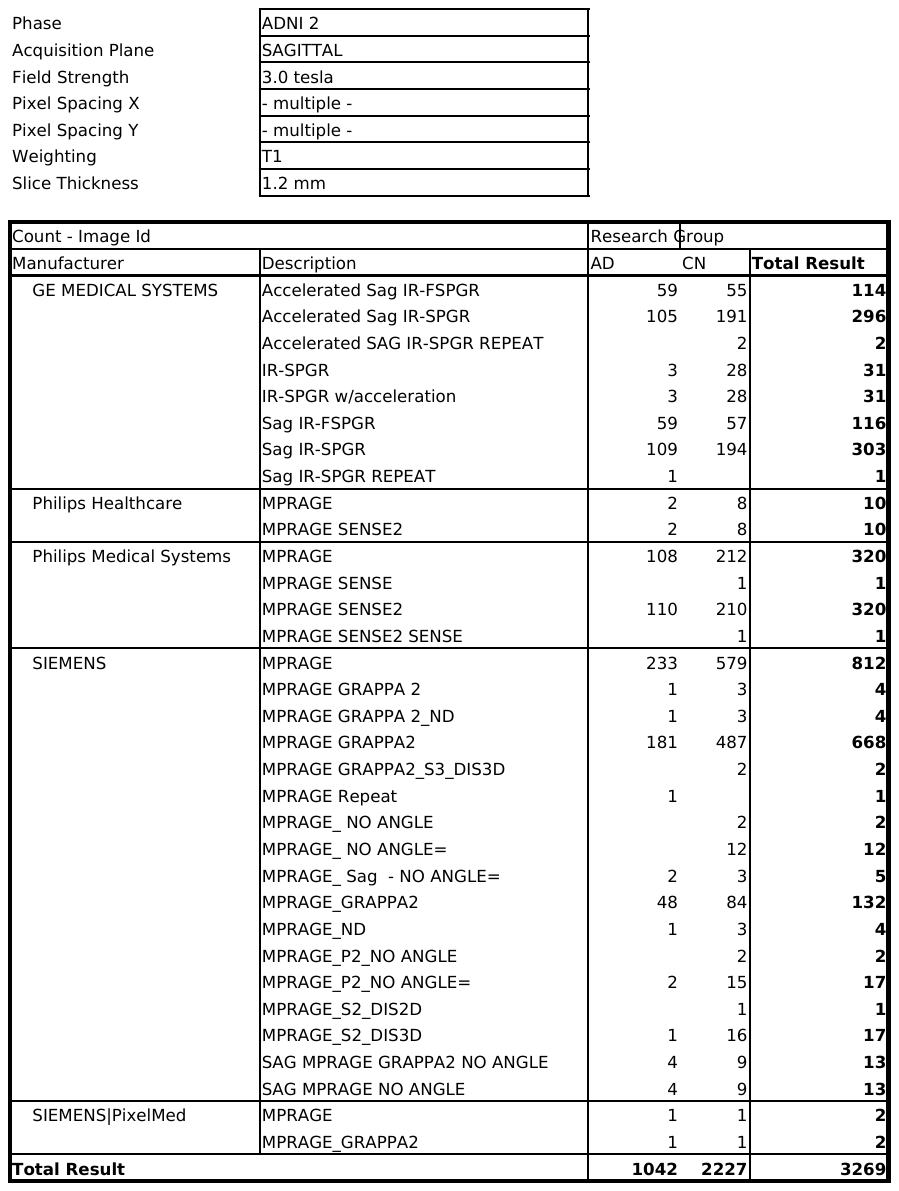}
  \par
}
\restoregeometry

\begin{figure}
    \centering
    \includegraphics[width=1.0\textwidth]{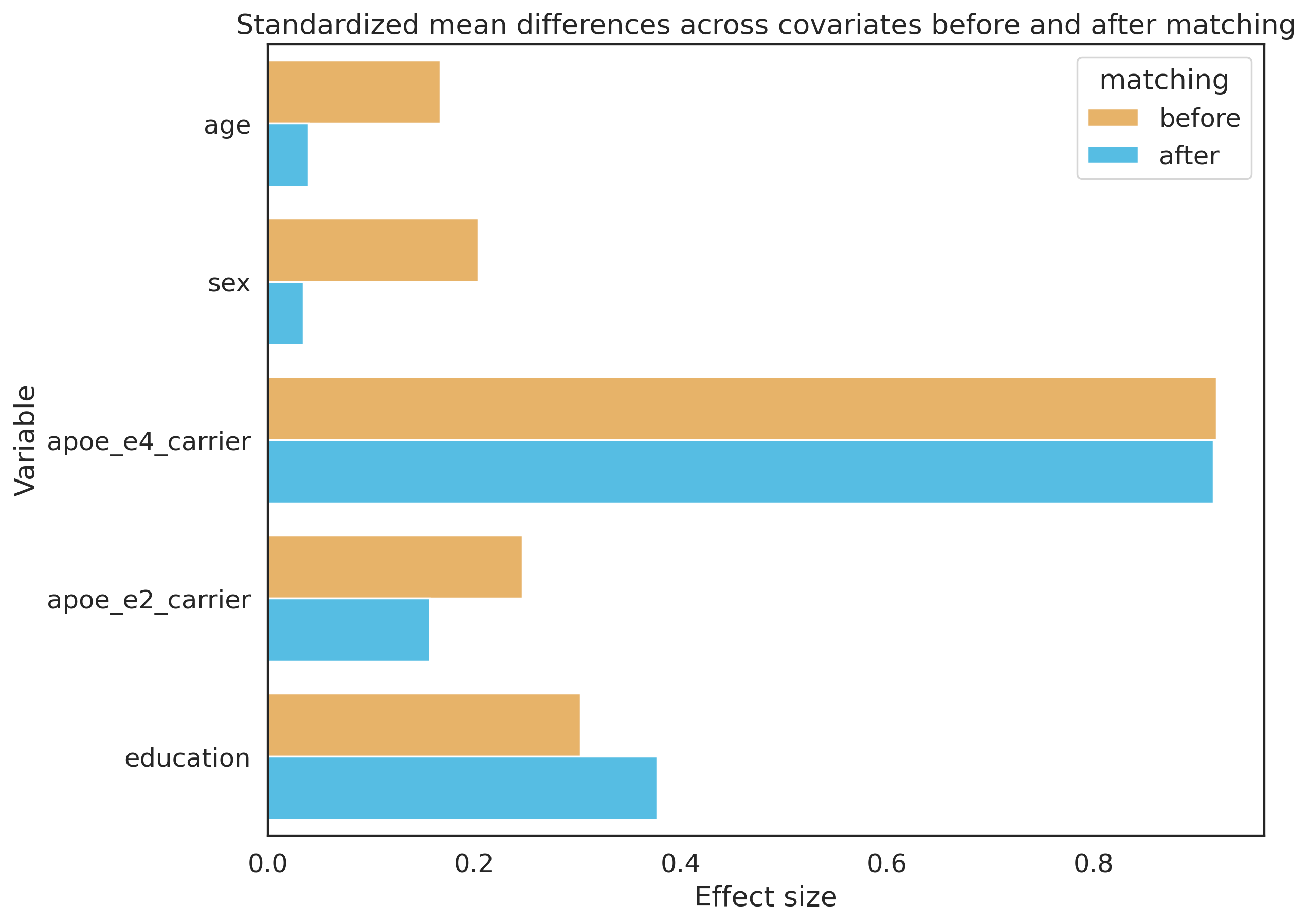}
    \caption{Standardized mean differences across covariates before and after matching. Only variables age and sex were used for propensity-logit-matching.}
    \label{fig:effect_sizes}
\end{figure}

Figure~\ref{fig:effect_sizes} shows standardized mean differences across covariates before and after propensity-logit-matching. APOE $\epsilon$4 carrier status, which has been associated with subtle regional gray matter and hippocampal volume differences even in cognitively normal individuals, was not included as a matching covariate. As the objective of this study was to assess preprocessing- and model-related biases rather than genetic or disease-risk effects, only age and sex were used for propensity matching. The resulting variation in APOE $\epsilon$4 distribution was therefore considered to reflect natural biological heterogeneity between groups and is therefore reflected in the effect size.

\begin{figure}
    \centering
    \includegraphics[width=1.0\textwidth]{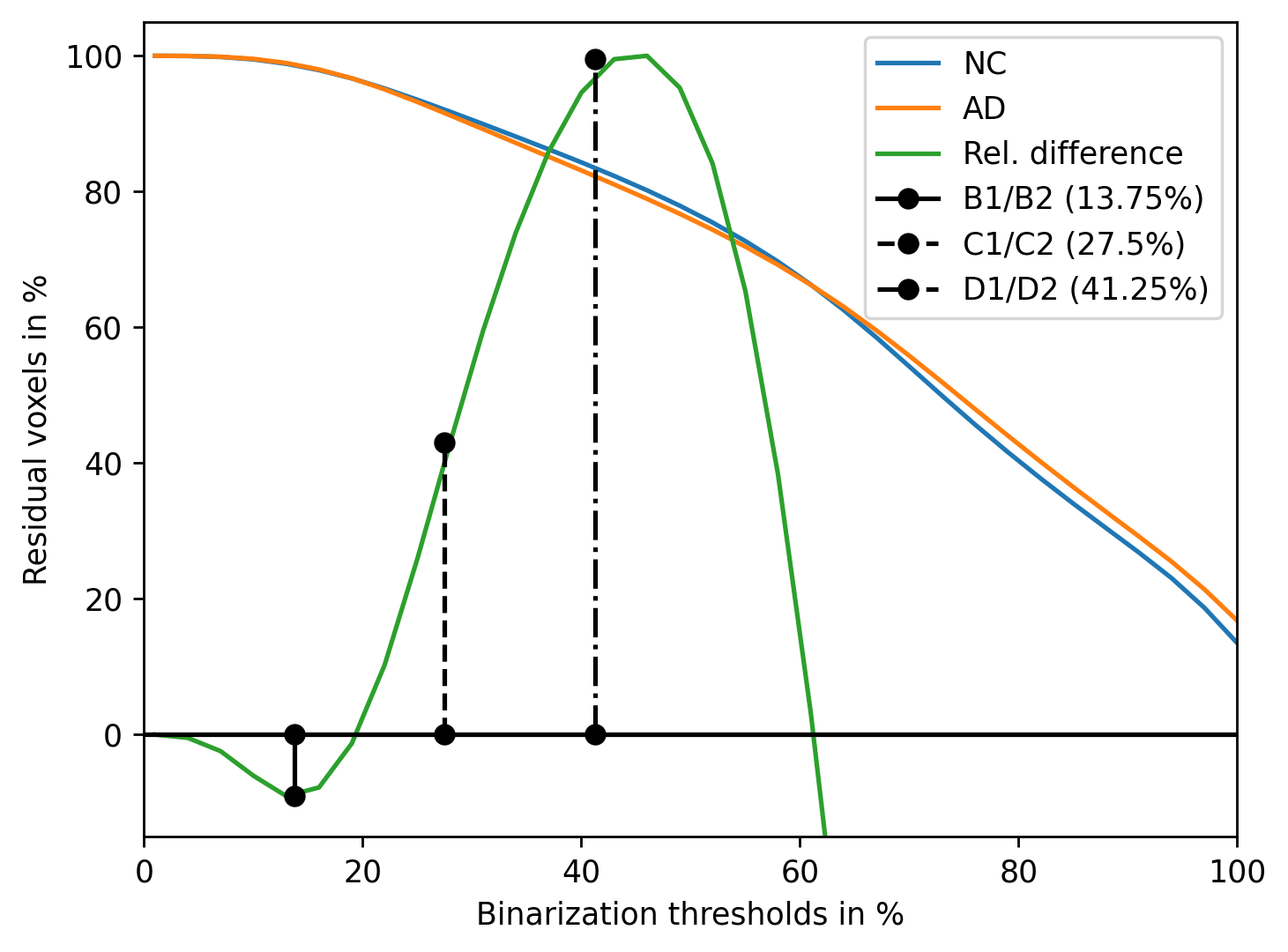}
    \caption{Residual voxels expressed as a percentage of the brain mask across binarization thresholds. Normalized group differences (NC vs. AD) are shown, with black lines indicating thresholds applied in the model setups.}
    \label{fig:group_divergence}
\end{figure}

\clearpage
Although the binarization thresholds in the model setup were chosen arbitrarily, Figure~\ref{fig:group_divergence} illustrates that they were selected to preserve meaningful atrophy patterns by comparing residual voxels with individual brain masks. Thresholds above 50\% were excluded, as they resulted in visually unrealistic images not identifiable as brains.

\section*{Supplementary Material 2}
\label{sec:sm_2}

\begin{figure}[H]
    \centering
    \includegraphics[width=0.75\textwidth]{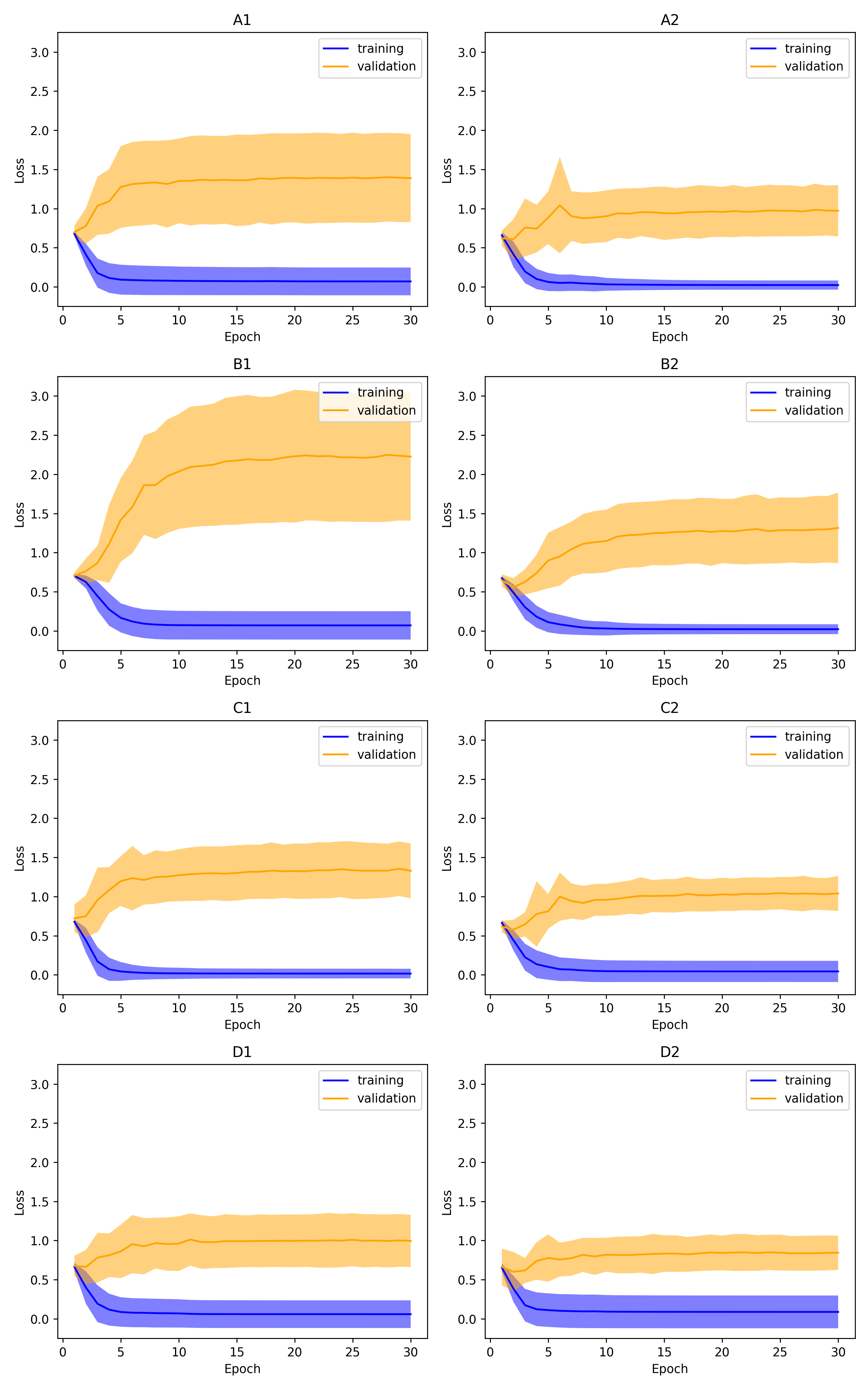}
    \caption{Mean and standard deviation of training (blue) and validation (orange) loss curves over epochs for all setups. Training loss decreases smoothly, while validation loss slightly rises before stabilizing. Both curves plateau after 10 epochs, indicating convergence. The gap between training and validation losses suggests minimal overfitting. The model architecture was optimized for setup A2.}
    \label{fig:loss_curves}
\end{figure}

\begin{figure}
    \centering
    \includegraphics[width=0.75\textwidth]{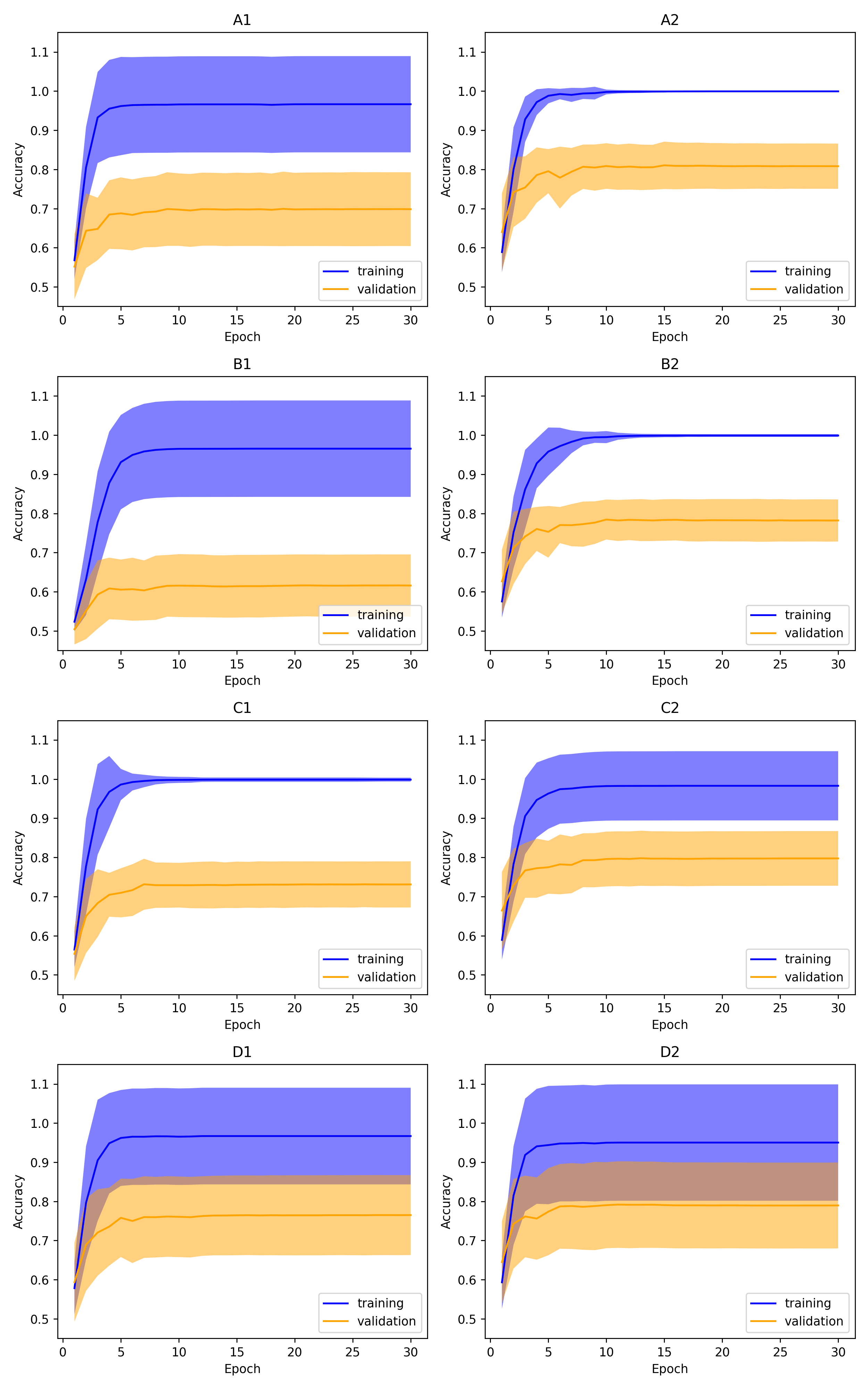}
    \caption{Mean and standard deviation of training (blue) and validation (orange) accuracy curves over epochs for all setups.}
    \label{fig:accuracy_curves}
\end{figure}

\clearpage
\section*{Supplementary Material 3}
\label{sec:sm_3}

\begin{longtable}{|l|r|r|r|r|r|}
    \caption{Multiple testing using two-sided p-values for multiple models and performance measures. Note. Comparison denotes models and measure; e.g., “A2-B1, acc” compares models A2 (skull-stripped, no-binarization) and B1 (binarized-13.75\%) with respect to the overall accuracy. $\Theta_1$ and $\Theta_2$ refers to the respective performance metrics. Adj. p-value is the discrete Bonferroni-Holm corrected p-value of the comparison. The table shows only the 122 significant differences of the multiple testing and the first non-significant difference. Remaining results are indicated with “(477 more)”.}
    \label{tab:statistical_comparison}\\
    \hline
    \textbf{Comparison} & \textbf{$\Theta_1$}            & \textbf{$\Theta_2$}            & \textbf{Sampling Index} & \textbf{Initial Weights Index} & \textbf{Adj. p-value}  \\ \hline\hline
    \endfirsthead
    \endhead
    A2-B1, acc          & 0.84                   & 0.55                   & 9                       & 2                              & 0.00000                \\ \hline
    A2-B1, acc          & 0.89                   & 0.58                   & 9                       & 3                              & 0.00000                \\ \hline
    A2-B1, acc          & 0.88                   & 0.61                   & 2                       & 3                              & 0.00000                \\ \hline
    A2-B1, acc          & 0.83                   & 0.54                   & 8                       & 1                              & 0.00000                \\ \hline
    A2-B1, spe          & 0.89                   & 0.50                   & 9                       & 2                              & 0.00000                \\ \hline
    A2-C1, spe          & 0.89                   & 0.52                   & 9                       & 2                              & 0.00000                \\ \hline
    A2-B1, acc          & 0.84                   & 0.62                   & 7                       & 3                              & 0.00000                \\ \hline
    A2-B1, acc          & 0.80                   & 0.53                   & 4                       & 1                              & 0.00000                \\ \hline
    A2-A1, acc          & 0.86                   & 0.68                   & 1                       & 2                              & 0.00000                \\ \hline
    A2-B1, spe          & 0.94                   & 0.64                   & 7                       & 3                              & 0.00000                \\ \hline
    A2-B1, acc          & 0.83                   & 0.58                   & 2                       & 2                              & 0.00000                \\ \hline
    A2-B1, acc          & 0.87                   & 0.62                   & 10                      & 2                              & 0.00000                \\ \hline
    A2-B1, acc          & 0.84                   & 0.59                   & 9                       & 1                              & 0.00000                \\ \hline
    A2-B1, acc          & 0.86                   & 0.63                   & 4                       & 2                              & 0.00000                \\ \hline
    A2-A1, sen          & 0.90                   & 0.57                   & 10                      & 2                              & 0.00000                \\ \hline
    A2-B1, sen          & 0.93                   & 0.63                   & 9                       & 3                              & 0.00000                \\ \hline
    A2-A1, sen          & 0.87                   & 0.60                   & 1                       & 2                              & 0.00000                \\ \hline
    A2-A1, spe          & 0.94                   & 0.69                   & 7                       & 3                              & 0.00000                \\ \hline
    A2-C1, acc          & 0.84                   & 0.61                   & 9                       & 2                              & 0.00000                \\ \hline
    A2-B1, sen          & 0.90                   & 0.59                   & 3                       & 3                              & 0.00000                \\ \hline
    A2-C1, spe          & 0.82                   & 0.52                   & 1                       & 3                              & 0.00000                \\ \hline
    A2-B1, acc          & 0.86                   & 0.68                   & 1                       & 2                              & 0.00000                \\ \hline
    A2-C1, acc          & 0.86                   & 0.67                   & 1                       & 2                              & 0.00000                \\ \hline
    A2-A1, acc          & 0.82                   & 0.60                   & 4                       & 3                              & 0.00000                \\ \hline
    A2-B1, acc          & 0.79                   & 0.57                   & 5                       & 3                              & 0.00000                \\ \hline
    A2-B1, spe          & 0.89                   & 0.61                   & 9                       & 1                              & 0.00000                \\ \hline
    A2-B1, acc          & 0.81                   & 0.61                   & 10                      & 1                              & 0.00000                \\ \hline
    A2-B1, sen          & 0.87                   & 0.58                   & 10                      & 3                              & 0.00000                \\ \hline
    A2-B1, sen          & 0.80                   & 0.47                   & 8                       & 1                              & 0.00000                \\ \hline
    A2-B1, sen          & 0.82                   & 0.51                   & 2                       & 3                              & 0.00000                \\ \hline
    A2-B1, sen          & 0.90                   & 0.60                   & 10                      & 2                              & 0.00001                \\ \hline
    A2-A1, acc          & 0.86                   & 0.67                   & 4                       & 2                              & 0.00001                \\ \hline
    A2-B2, acc          & 0.89                   & 0.73                   & 9                       & 3                              & 0.00001                \\ \hline
    A2-B1, acc          & 0.81                   & 0.59                   & 8                       & 2                              & 0.00001                \\ \hline
    A2-B1, spe          & 0.85                   & 0.52                   & 9                       & 3                              & 0.00001                \\ \hline
    A2-C1, acc          & 0.83                   & 0.64                   & 8                       & 1                              & 0.00001                \\ \hline
    A2-A1, acc          & 0.87                   & 0.70                   & 10                      & 2                              & 0.00001                \\ \hline
    A2-B1, spe          & 0.94                   & 0.72                   & 2                       & 3                              & 0.00001                \\ \hline
    A2-B1, spe          & 0.90                   & 0.62                   & 4                       & 2                              & 0.00001                \\ \hline
    A2-B1, acc          & 0.81                   & 0.63                   & 3                       & 1                              & 0.00002                \\ \hline
    A2-C1, acc          & 0.81                   & 0.61                   & 8                       & 2                              & 0.00002                \\ \hline
    A2-B1, acc          & 0.76                   & 0.56                   & 10                      & 3                              & 0.00003                \\ \hline
    A2-D2, spe          & 0.85                   & 0.66                   & 10                      & 2                              & 0.00003                \\ \hline
    A2-B1, spe          & 0.87                   & 0.61                   & 4                       & 3                              & 0.00004                \\ \hline
    A2-C1, sen          & 0.90                   & 0.64                   & 10                      & 2                              & 0.00004                \\ \hline
    A2-B2, spe          & 0.66                   & 0.89                   & 10                      & 3                              & 0.00006                \\ \hline
    A2-D1, sen          & 0.82                   & 0.53                   & 4                       & 2                              & 0.00006                \\ \hline
    A2-A1, acc          & 0.81                   & 0.62                   & 8                       & 2                              & 0.00007                \\ \hline
    A2-C1, sen          & 0.80                   & 0.55                   & 8                       & 1                              & 0.00008                \\ \hline
    A2-C1, acc          & 0.87                   & 0.71                   & 10                      & 2                              & 0.00008                \\ \hline
    A2-B1, acc          & 0.82                   & 0.64                   & 4                       & 3                              & 0.00008                \\ \hline
    A2-D1, acc          & 0.86                   & 0.71                   & 1                       & 2                              & 0.00009                \\ \hline
    A2-B1, spe          & 0.83                   & 0.52                   & 4                       & 1                              & 0.00009                \\ \hline
    A2-A1, spe          & 0.90                   & 0.64                   & 4                       & 2                              & 0.00011                \\ \hline
    A2-A1, sen          & 0.94                   & 0.74                   & 5                       & 2                              & 0.00013                \\ \hline
    A2-B1, spe          & 0.86                   & 0.61                   & 8                       & 1                              & 0.00014                \\ \hline
    A2-B1, acc          & 0.84                   & 0.68                   & 1                       & 1                              & 0.00015                \\ \hline
    A2-B1, sen          & 0.81                   & 0.54                   & 2                       & 2                              & 0.00024                \\ \hline
    A2-A1, acc          & 0.88                   & 0.75                   & 2                       & 3                              & 0.00025                \\ \hline
    A2-A1, sen          & 0.75                   & 0.47                   & 8                       & 2                              & 0.00025                \\ \hline
    A2-A1, acc          & 0.81                   & 0.66                   & 3                       & 1                              & 0.00028                \\ \hline
    A2-C1, acc          & 0.81                   & 0.66                   & 1                       & 3                              & 0.00031                \\ \hline
    A2-D1, spe          & 0.66                   & 0.87                   & 10                      & 3                              & 0.00032                \\ \hline
    A2-C1, sen          & 0.87                   & 0.64                   & 1                       & 2                              & 0.00034                \\ \hline
    A2-B1, sen          & 0.78                   & 0.54                   & 4                       & 1                              & 0.00039                \\ \hline
    A2-A1, spe          & 0.87                   & 0.66                   & 4                       & 3                              & 0.00042                \\ \hline
    A2-B1, acc          & 0.82                   & 0.67                   & 7                       & 2                              & 0.00052                \\ \hline
    A2-B1, spe          & 0.85                   & 0.62                   & 2                       & 2                              & 0.00065                \\ \hline
    A2-B1, spe          & 0.92                   & 0.71                   & 7                       & 1                              & 0.00105                \\ \hline
    A2-D1, sen          & 0.90                   & 0.66                   & 10                      & 2                              & 0.00104                \\ \hline
    A2-B1, spe          & 0.83                   & 0.60                   & 3                       & 1                              & 0.00108                \\ \hline
    A2-C1, acc          & 0.82                   & 0.69                   & 7                       & 2                              & 0.00162                \\ \hline
    A2-C1, acc          & 0.89                   & 0.75                   & 9                       & 3                              & 0.00161                \\ \hline
    A2-B2, sen          & 0.82                   & 0.63                   & 4                       & 2                              & 0.00224                \\ \hline
    A2-D2, acc          & 0.79                   & 0.89                   & 5                       & 3                              & 0.00224                \\ \hline
    A2-B1, sen          & 0.75                   & 0.48                   & 8                       & 2                              & 0.00235                \\ \hline
    A2-D2, sen          & 0.78                   & 0.58                   & 4                       & 1                              & 0.00278                \\ \hline
    A2-B2, acc          & 0.86                   & 0.74                   & 4                       & 2                              & 0.00281                \\ \hline
    A2-B1, acc          & 0.80                   & 0.66                   & 7                       & 1                              & 0.00285                \\ \hline
    A2-B2, spe          & 0.85                   & 0.64                   & 9                       & 3                              & 0.00289                \\ \hline
    A2-D2, spe          & 0.65                   & 0.81                   & 5                       & 3                              & 0.00294                \\ \hline
    A2-A1, spe          & 0.89                   & 0.71                   & 9                       & 2                              & 0.00385                \\ \hline
    A2-A1, acc          & 0.84                   & 0.72                   & 1                       & 1                              & 0.00471                \\ \hline
    A2-D1, acc          & 0.86                   & 0.71                   & 4                       & 2                              & 0.00612                \\ \hline
    A2-B1, spe          & 0.80                   & 0.59                   & 10                      & 1                              & 0.00760                \\ \hline
    A2-D1, sen          & 0.87                   & 0.66                   & 1                       & 2                              & 0.00759                \\ \hline
    A2-B1, sen          & 0.82                   & 0.60                   & 6                       & 1                              & 0.00776                \\ \hline
    A2-B2, spe          & 0.94                   & 0.76                   & 7                       & 3                              & 0.00774                \\ \hline
    A2-C1, spe          & 0.88                   & 0.65                   & 8                       & 2                              & 0.00779                \\ \hline
    A2-A1, sen          & 0.81                   & 0.62                   & 2                       & 1                              & 0.00786                \\ \hline
    A2-A1, acc          & 0.82                   & 0.69                   & 7                       & 2                              & 0.00822                \\ \hline
    A2-B1, spe          & 0.65                   & 0.39                   & 5                       & 3                              & 0.00861                \\ \hline
    A2-A1, acc          & 0.89                   & 0.77                   & 9                       & 3                              & 0.00975                \\ \hline
    A2-B2, sen          & 0.75                   & 0.88                   & 7                       & 3                              & 0.00985                \\ \hline
    A2-B1, sen          & 0.79                   & 0.59                   & 9                       & 2                              & 0.01137                \\ \hline
    A2-C1, sen          & 0.81                   & 0.61                   & 2                       & 2                              & 0.01135                \\ \hline
    A2-A1, sen          & 0.79                   & 0.59                   & 1                       & 3                              & 0.01203                \\ \hline
    A2-B1, spe          & 0.85                   & 0.67                   & 1                       & 2                              & 0.01280                \\ \hline
    A2-A1, acc          & 0.80                   & 0.69                   & 7                       & 1                              & 0.01469                \\ \hline
    A2-B1, spe          & 0.85                   & 0.63                   & 10                      & 2                              & 0.01656                \\ \hline
    A2-A1, acc          & 0.79                   & 0.68                   & 2                       & 1                              & 0.01664                \\ \hline
    A2-A1, spe          & 0.94                   & 0.79                   & 2                       & 3                              & 0.01694                \\ \hline
    A2-D1, spe          & 0.65                   & 0.85                   & 5                       & 3                              & 0.01915                \\ \hline
    A2-C1, sen          & 0.93                   & 0.77                   & 9                       & 3                              & 0.01916                \\ \hline
    A2-C1, acc          & 0.81                   & 0.68                   & 10                      & 1                              & 0.01927                \\ \hline
    A2-A1, spe          & 0.90                   & 0.72                   & 7                       & 2                              & 0.02008                \\ \hline
    A2-C1, sen          & 0.90                   & 0.73                   & 3                       & 3                              & 0.02038                \\ \hline
    A2-B1, sen          & 0.87                   & 0.69                   & 1                       & 2                              & 0.02084                \\ \hline
    A2-C1, acc          & 0.83                   & 0.70                   & 2                       & 2                              & 0.02126                \\ \hline
    A2-B1, sen          & 0.79                   & 0.57                   & 9                       & 1                              & 0.02497                \\ \hline
    A2-B1, spe          & 0.88                   & 0.69                   & 1                       & 1                              & 0.02553                \\ \hline
    A2-C1, acc          & 0.86                   & 0.72                   & 4                       & 2                              & 0.02597                \\ \hline
    A2-A1, acc          & 0.83                   & 0.71                   & 8                       & 1                              & 0.02820                \\ \hline
    A2-B2, spe          & 0.66                   & 0.81                   & 3                       & 3                              & 0.02896                \\ \hline
    A2-B1, spe          & 0.90                   & 0.72                   & 7                       & 2                              & 0.02998                \\ \hline
    A2-C2, spe          & 0.66                   & 0.81                   & 3                       & 3                              & 0.03210                \\ \hline
    A2-A1, sen          & 0.82                   & 0.64                   & 10                      & 1                              & 0.03369                \\ \hline
    A2-B1, sen          & 0.93                   & 0.75                   & 5                       & 3                              & 0.03393                \\ \hline
    A2-A1, acc          & 0.84                   & 0.74                   & 7                       & 3                              & 0.03642                \\ \hline
    A2-B1, acc          & 0.77                   & 0.63                   & 3                       & 2                              & 0.03756                \\ \hline
    A2-B1, acc          & 0.79                   & 0.66                   & 2                       & 1                              & 0.03768                \\ \hline
    A2-C1, acc          & 0.81                   & 0.69                   & 3                       & 1                              & 0.03935                \\ \hline
    A2-C1, spe          & 0.89                   & 0.72                   & 9                       & 1                              & 0.05069                \\ \hline
    (477 more)          & \multicolumn{1}{c|}{$\vdots$} & \multicolumn{1}{c|}{$\vdots$} & \multicolumn{1}{c|}{$\vdots$}  & \multicolumn{1}{c|}{$\vdots$}         & \multicolumn{1}{c|}{$\vdots$} \\ \hline
\end{longtable}

\clearpage
\section*{Supplementary Material 4}
\label{sec:sm_4}

\begin{figure}
    \centering
    \includegraphics[width=0.7\textwidth]{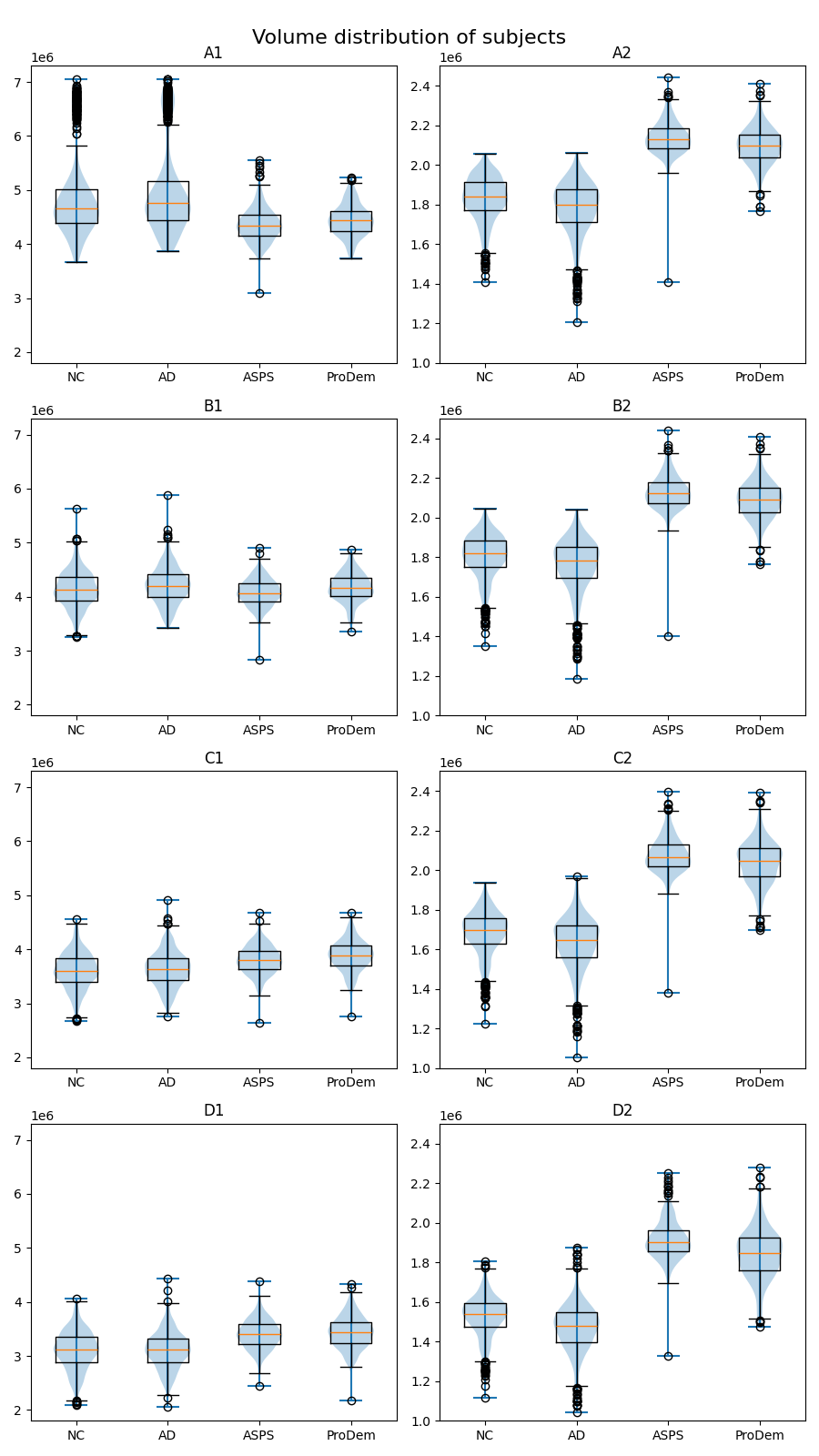}
    \caption{Comparison of total voxel count distributions between NC and AD groups from ADNI and our local cohorts ASPS (as controls) and ProDem (as patients) across the eight image setups. The left column shows (A1) aligned T1w MRIs and corresponding binarized images using thresholds of (B1) 13.75\%, (C1) 27.50\%, and (D1) 41.25\%. The right column displays the skull-stripped versions (A2, B2, C2, D2). Distributions are similar in the left column, while skull-stripping introduces a group-dependent offset in voxel counts. For A1 and A2, a low binarization threshold (5\% of white matter peak) was used to suppress background voxels.}
    \label{fig:volume_distributions}
\end{figure}

Table~\ref{tab:demographics_our} presents the demographic characteristics of our local, non-public datasets at baseline. These datasets were age-matched to the subset of ADNI data used for model training. The corresponding model performance on these local datasets is summarized in Table~\ref{tab:performance_our}.
\\~\\
The reference model (A2) exhibits an average performance decline of approximately 9\% compared to its test performance on ADNI data. Models B2 and C2 perform near chance level, while model D2 shows a performance drop of about 13\%. Notably, model A1 achieves nearly identical average performance on both the ADNI test set and our local datasets. Although models B1 and C1 underperform on the local data, the decline observed for model D1 is more moderate (approximately 8\%).
\\~\\
Importantly, the image preprocessing steps were identical for both ADNI and local datasets. To investigate potential causes of the performance discrepancies, we examined volume distributions across cohorts (Figure~\ref{fig:volume_distributions}). When skull-stripping was applied, a systematic offset was observed between the ADNI data and our local datasets. This offset may account for the pronounced performance decline in models B2 and C2, which rely primarily on volumetric features, as well as in model D2, which retains more structural information. The observed volume shift also suggests a potential contribution of skull-stripping to the diminished performance of model A2.

\begin{table}
\caption{Summary of subject demographics at baseline for our local, non-public datasets. Note. Values are presented as mean ± SD [range]. M: male, F: female, MMSE: mini-mental state examination, CDR: global clinical dementia rating, APOE: Apolipoprotein E status, Education in years, ASPS = Austrian Stroke Prevention Study \cite{schmidt_progression_2003} used as group normal control, ProDem = Prospective Dementia Registry Austria \cite{damulina_cross-sectional_2020} used as group Alzheimer's disease, n/a: no value available.
*CDR scores for ASPS were acquired at followup, no baseline is available.}
\label{tab:demographics_our}
\resizebox{\textwidth}{!}{%
\begin{tabular}{|l|l|l|l|l|l|l|l|l|}
\hline
       & \textbf{Subjects} & \textbf{Images} & \textbf{Age}                                                        & \textbf{Gender}                                        & \textbf{MMSE}                                                                & \textbf{CDR}                                                                   & \textbf{APOE}                                                                                                               & \textbf{Education}                                                          \\ \hline \hline
ASPS   & 304               & 401             & \begin{tabular}[c]{@{}l@{}}71.2±6.3\\ {[}55.7, 87.2{]}\end{tabular} & \begin{tabular}[c]{@{}l@{}}119 M/\\ 185 F\end{tabular} & \begin{tabular}[c]{@{}l@{}}27.9±1.6\\ {[}22.0, 30.0{]}\\ n/a: 0\end{tabular} & \begin{tabular}[c]{@{}l@{}}0.5: 21;\\ n/a: 283*\end{tabular}                   & \begin{tabular}[c]{@{}l@{}}$\epsilon$2/$\epsilon$2: 2;\\ $\epsilon$2/$\epsilon$3: 33;\\ $\epsilon$2/$\epsilon$4: 1;\\ $\epsilon$3/$\epsilon$3: 209;\\ $\epsilon$3/$\epsilon$4: 43;\\ $\epsilon$4/$\epsilon$4: 4;\\ n/a: 12\end{tabular} & \begin{tabular}[c]{@{}l@{}}11.3±2.8\\ {[}9.0, 18.0{]}\\ n/a: 0\end{tabular} \\ \hline
ProDem & 178               & 358             & \begin{tabular}[c]{@{}l@{}}73.3±8.0\\ {[}55.7, 90.7{]}\end{tabular} & \begin{tabular}[c]{@{}l@{}}75 M/\\ 103 F\end{tabular}  & \begin{tabular}[c]{@{}l@{}}21.7±4.2\\ {[}7.0, 29.0{]}\\ n/a: 1\end{tabular}  & \begin{tabular}[c]{@{}l@{}}0.5: 90;\\ 1.0: 77;\\ 2.0: 6;\\ n/a: 5\end{tabular} & \begin{tabular}[c]{@{}l@{}}$\epsilon$2/$\epsilon$2: 1;\\ $\epsilon$2/$\epsilon$3: 7;\\ $\epsilon$2/$\epsilon$4: 7;\\ $\epsilon$3/$\epsilon$3: 87;\\ $\epsilon$3/$\epsilon$4: 67;\\ $\epsilon$4/$\epsilon$4: 8;\\ n/a: 1\end{tabular}    & \begin{tabular}[c]{@{}l@{}}11.0±2.7\\ {[}9.0, 18.0{]}\\ n/a: 1\end{tabular} \\ \hline
\end{tabular}
}
\end{table}

\begin{table}
\caption{Summary of performance metrics of all configurations on local datasets (ASPS \cite{schmidt_progression_2003}, ProDem \cite{damulina_cross-sectional_2020}). Note. AUC = area under receiver operating characteristics curve.
Column Id refers to preprocessing defined in Figure~\ref{fig:example}.
Values between [ and ] show the 95\% confidence interval.
}
\label{tab:performance_our}
\resizebox{\textwidth}{!}{%
\begin{tabular}{|l|l|l|l|l|l|l|}
\hline
\textbf{Input images}               & \textbf{Id} & \textbf{Binarizer} & \textbf{Accuracy}                                                                      & \textbf{Sensitivity}                                                                    & \textbf{Specificity}                                                                   & \textbf{AUC}                                                                   \\ \hline \hline
\multirow{7}{*}{Aligned T1w}        & A1          & None               & \begin{tabular}[c]{@{}l@{}}72.31±2.11\%\\ {[}68.87\%, 75.75\%{]}\end{tabular}          & \begin{tabular}[c]{@{}l@{}}58.75±7.16\%\\ {[}47.32\%, 71.97\%{]}\end{tabular}           & \textbf{\begin{tabular}[c]{@{}l@{}}84.41±5.15\%\\ {[}75.65\%, 93.27\%{]}\end{tabular}} & \begin{tabular}[c]{@{}l@{}}0.72±0.022 \\ {[}0.68, 0.75{]}\end{tabular}         \\ \cline{2-7} 
                                    & B1          & 13.75\%            & \begin{tabular}[c]{@{}l@{}}52.53±3.22\%\\ {[}47.52\%, 58.54\%{]}\end{tabular}          & \begin{tabular}[c]{@{}l@{}}61.27±14.33\%\\ {[}35.22\%, 84.89\%{]}\end{tabular}          & \begin{tabular}[c]{@{}l@{}}44.72±16.19\%\\ {[}18.79\%, 71.02\%{]}\end{tabular}         & \begin{tabular}[c]{@{}l@{}}0.53±0.028\\ {[}0.49, 0.58{]}\end{tabular}          \\ \cline{2-7} 
                                    & C1          & 27.50\%            & \begin{tabular}[c]{@{}l@{}}51.93±3.91\%\\ {[}46.25\%, 59.30\%{]}\end{tabular}          & \begin{tabular}[c]{@{}l@{}}70.25±16.62\%\\ {[}38.67\%, 93.88\%{]}\end{tabular}          & \begin{tabular}[c]{@{}l@{}}35.58±20.35\%\\ {[}4.49\%, 69.94\%{]}\end{tabular}          & \begin{tabular}[c]{@{}l@{}}0.53±0.032\\ {[}0.48, 0.59{]}\end{tabular}          \\ \cline{2-7} 
                                    & D1          & 41.25\%            & \begin{tabular}[c]{@{}l@{}}69.50±3.43\%\\ {[}61.95\%, 74.59\%{]}\end{tabular}          & \begin{tabular}[c]{@{}l@{}}63.77±10.27\%\\ {[}46.96\%, 84.39\%{]}\end{tabular}          & \begin{tabular}[c]{@{}l@{}}74.63±11.62\%\\ {[}51.06\%, 91.17\%{]}\end{tabular}         & \begin{tabular}[c]{@{}l@{}}0.69±0.033\\ {[}0.62, 0.74{]}\end{tabular}          \\ \hline
\multirow{7}{*}{Skull-stripped T1w} & A2          & None               & \textbf{\begin{tabular}[c]{@{}l@{}}72.90±4.37\%\\ {[}63.21\%, 79.78\%{]}\end{tabular}} & \textbf{\begin{tabular}[c]{@{}l@{}}70.66±11.78\%\\ {[}50.89\%, 91.09\%{]}\end{tabular}} & \begin{tabular}[c]{@{}l@{}}74.90±16.19\%\\ {[}39.84\%, 94.51\%{]}\end{tabular}         & \textbf{\begin{tabular}[c]{@{}l@{}}0.73±0.039\\ {[}0.64, 0.79{]}\end{tabular}} \\ \cline{2-7} 
                                    & B2          & 13.75\%            & \begin{tabular}[c]{@{}l@{}}58.47±5.55\%\\ {[}47.93\%, 68.32\%{]}\end{tabular}          & \begin{tabular}[c]{@{}l@{}}54.72±17.37\%\\ {[}24.88\%, 80.33\%{]}\end{tabular}          & \begin{tabular}[c]{@{}l@{}}61.81±23.95\%\\ {[}22.17\%, 93.90\%{]}\end{tabular}         & \begin{tabular}[c]{@{}l@{}}0.58±0.047\\ {[}0.49, 0.67{]}\end{tabular}          \\ \cline{2-7} 
                                    & C2          & 27.50\%            & \begin{tabular}[c]{@{}l@{}}52.20±5.63\%\\ {[}44.60\%, 63.69\%{]}\end{tabular}          & \begin{tabular}[c]{@{}l@{}}69.76±16.61\%\\ {[}33.91\%, 96.01\%{]}\end{tabular}          & \begin{tabular}[c]{@{}l@{}}36.51±23.81\%\\ {[}0.42\%, 82.17\%{]}\end{tabular}          & \begin{tabular}[c]{@{}l@{}}0.53±0.047\\ {[}0.46, 0.62{]}\end{tabular}          \\ \cline{2-7} 
                                    & D2          & 41.25\%            & \begin{tabular}[c]{@{}l@{}}68.11±5.33\%\\ {[}57.33\%, 75.11\%{]}\end{tabular}          & \begin{tabular}[c]{@{}l@{}}62.96±13.70\%\\ {[}36.56\%, 84.71\%{]}\end{tabular}          & \begin{tabular}[c]{@{}l@{}}72.71±17.65\%\\ {[}38.24\%, 95.40\%{]}\end{tabular}         & \begin{tabular}[c]{@{}l@{}}0.68±0.049\\ {[}0.58, 0.75{]}\end{tabular}          \\ \hline
\end{tabular}
}
\end{table}

\clearpage
\section*{Supplementary Material 5}
\label{sec:sm_5}

\subsection*{Introduction to spectral relevance analysis}

Spectral relevance analysis (SpRAy) enables efficient exploration of classifier behavior across large datasets by applying spectral clustering to inputs and heatmaps. This technique identifies common and atypical decision-making patterns, highlighting image features that may or may not reflect clinically relevant concepts. SpRAy is useful for uncovering unexpected or artifact-driven classifier behaviors, similar to the Clever Hans effect found in \cite{lapuschkin_unmasking_2019}.
\\~\\
The SpRAy process implemented for this study involves five steps:
\begin{enumerate}
    \item 
Compute relevance maps using LRP to identify focus areas for classification.
    \item
Downsample the inputs and the heatmaps to 2 mm isotropic resolution for efficient analysis.
    \item
Perform spectral clustering to group similar image or relevance patterns.
    \item
Identify clusters with highest eigenvalue gap, indicating well-separated groups, and compute mean heatmaps for groups.
\end{enumerate}
Visualize the clusters using t-distributed stochastic neighbor embedding (t-SNE) \cite{maaten_visualizing_2008}, which aids in interpreting the results and understanding the relationship between clusters.

\subsection*{Results of spectral relevance analysis}

We visualized the clustering of inputs and heatmaps using t-SNE, initialized with the normalized, symmetric, and positive semi-definite Laplacian matrix derived from the spectral clustering affinity matrix. Figure~\ref{fig:clustering} illustrates the grouping of inputs and corresponding heatmaps for the reference model (skull-stripped, no binarization, A2), the 27.5\% binarized skull-stripped model (C2), and the 41.25\% binarized model without skull-stripping (D1). Group mean heatmaps, based on spectral clustering groupings, are presented in Figure~\ref{fig:group_heatmaps}.
\\~\\
The analysis of the eigenvalues and the eigenvalue gaps of the Laplacians and the t-SNE visualization of input and heatmap groupings for all eight models are given in Figures~\ref{fig:appendix_1}, ~\ref{fig:appendix_2}, ~\ref{fig:appendix_3}, and ~\ref{fig:appendix_4}. Additionally, Figures~\ref{fig:binADNI_supplemental_2_figure_1} and ~\ref{fig:binADNI_supplemental_2_figure_2} present the analysis of misclassified samples and the corresponding heatmaps for models A2 and C2 using SpRAy.

\subsection*{Discussion of spectral relevance analysis}

\begin{figure}
    \includegraphics[width=\textwidth]{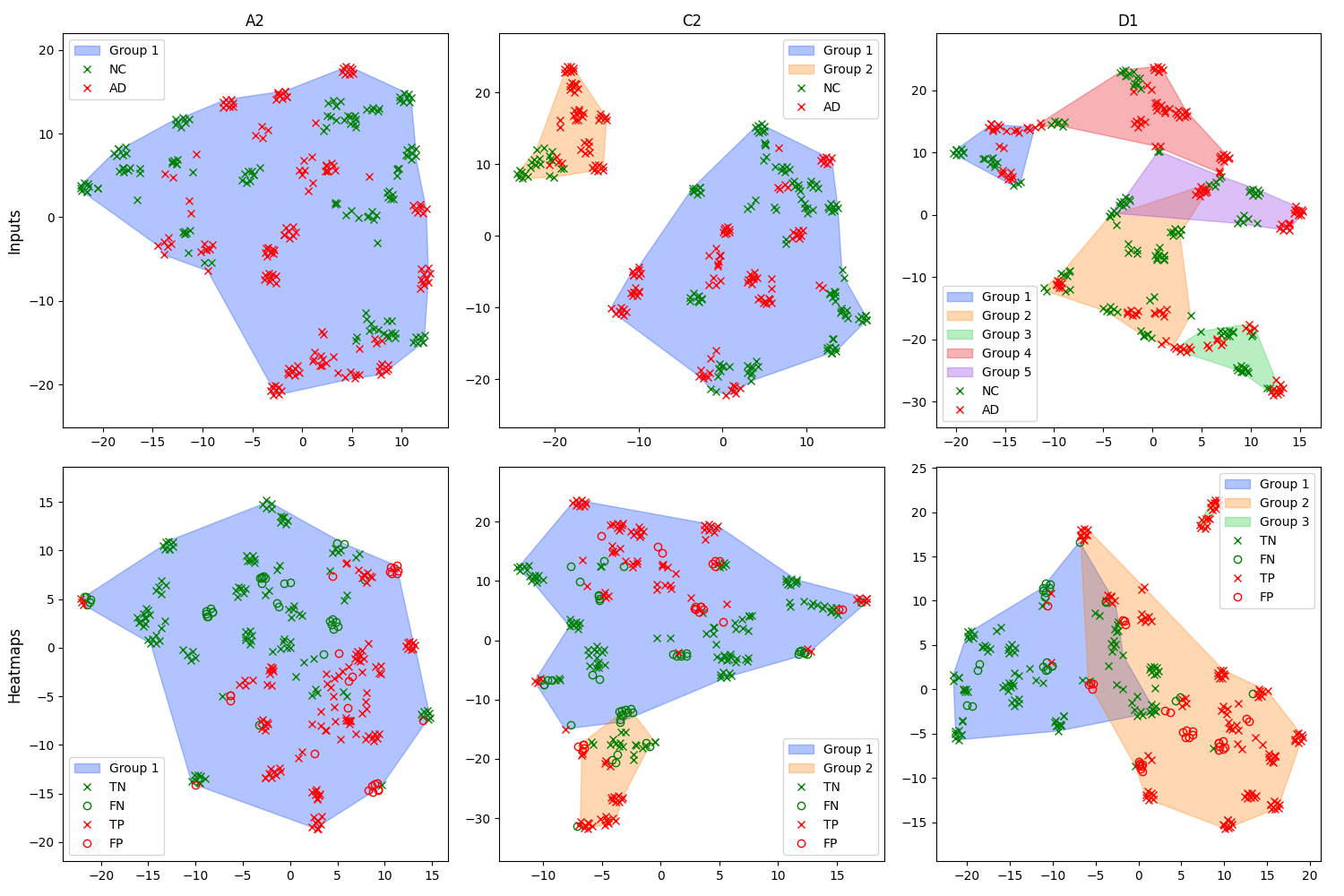}
    \caption{t-SNE visualization of inputs (row 1) and heatmaps (row 2) for the reference model (A2), the 27.5\% binarization model (C2), and the 41.25\% binarization model without skull-stripping (D1). Input data points are labeled by group (NC or AD), while heatmap points are categorized by confusion matrix outcomes (TN, FN, TP, FP). Only the heatmaps of model D1 exhibit clustering that aligns with the subject groups (NC vs. AD). Note. NC: normal control; AD: Alzheimer’s disease; TN: true negative; FN: false negative; TP: true positive; FP: false positive.}
    \label{fig:clustering}
\end{figure}

\begin{figure}
    \centering
    \includegraphics[width=0.85\textwidth]{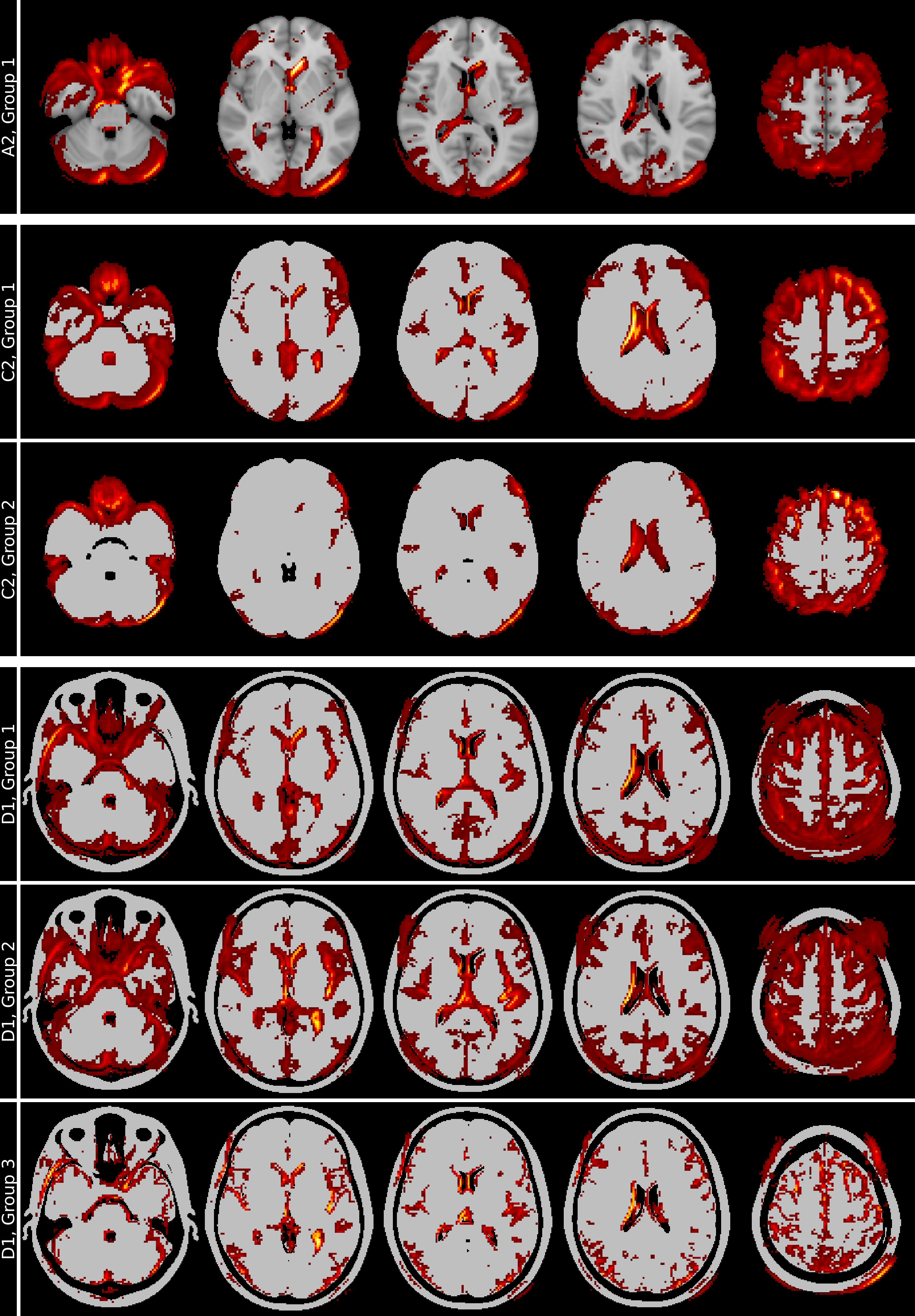}
    \caption{Mean heatmaps for the groups identified using the individual heatmaps and spectral clustering for models A2, C2, and D1. Model D1 (41.25\% binarization without skull-stripping) shows a clearer separation between AD (Group 2) and NC (Group 1) than A2 and C2. Group 2 in D1 highlights the left insular cortex, suggesting a more structured relevance pattern. Group 3, similar to Group 2, also shows relevance in the skull, indicating an unexpected decision strategy for AD classification. Images are shown in standard-radiological view, causing the left and right side of the brain to be flipped.}
    \label{fig:group_heatmaps}
\end{figure}

\begin{figure}
    \includegraphics[width=\textwidth]{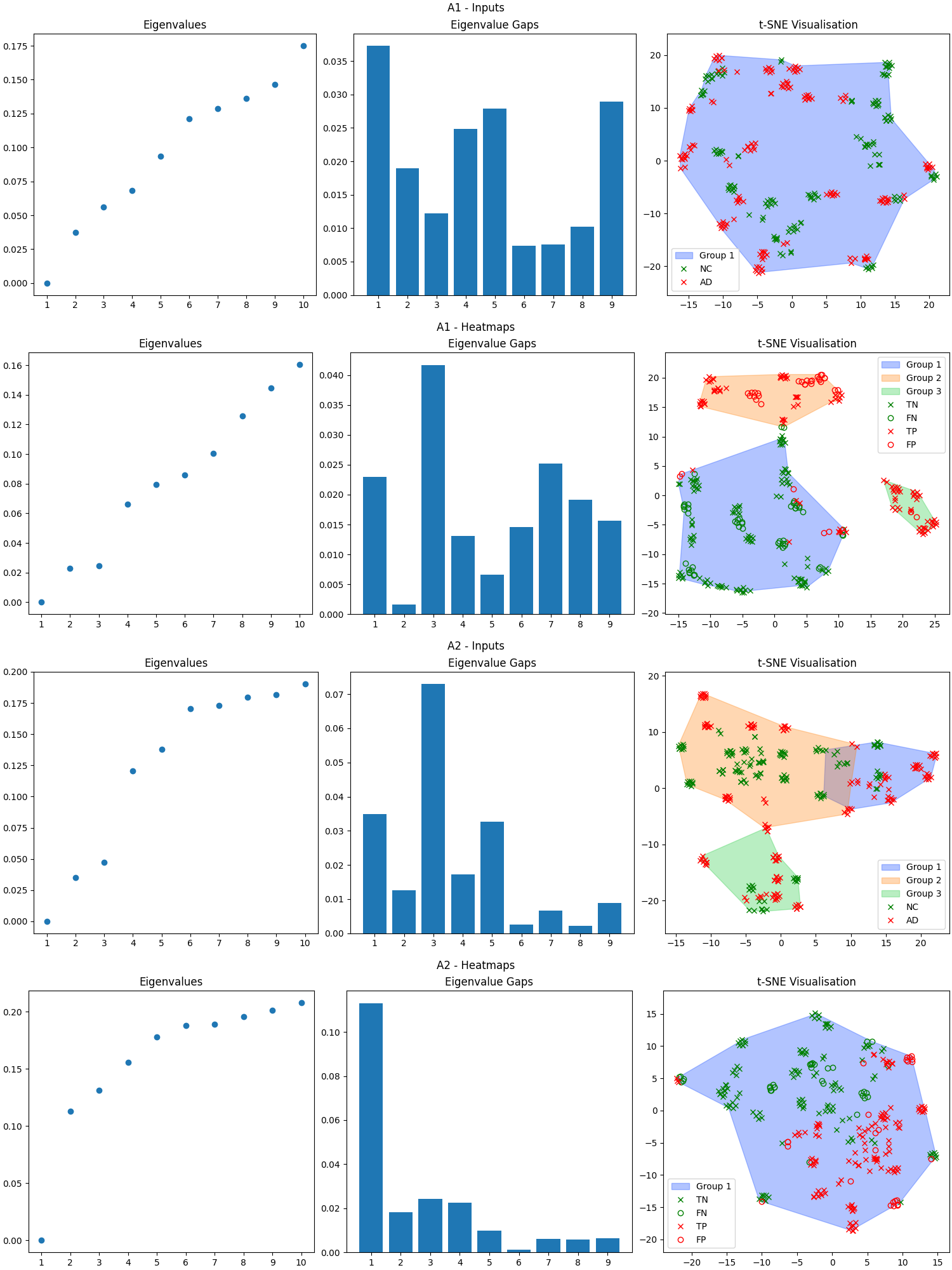}
    \caption{Eigenvalues (column 1), the respective eigenvalue gaps (column 2) of the Laplacian matrix and the and t-SNE visualization (column 3) of input (rows 1 and 3) and heatmap groupings (rows 2 and 4) for models A1 (aligned T1w MRI) and A2 (skull-stripped T1w MRI).}
    \label{fig:appendix_1}
\end{figure}

\begin{figure}
    \centering
    \includegraphics[width=\textwidth]{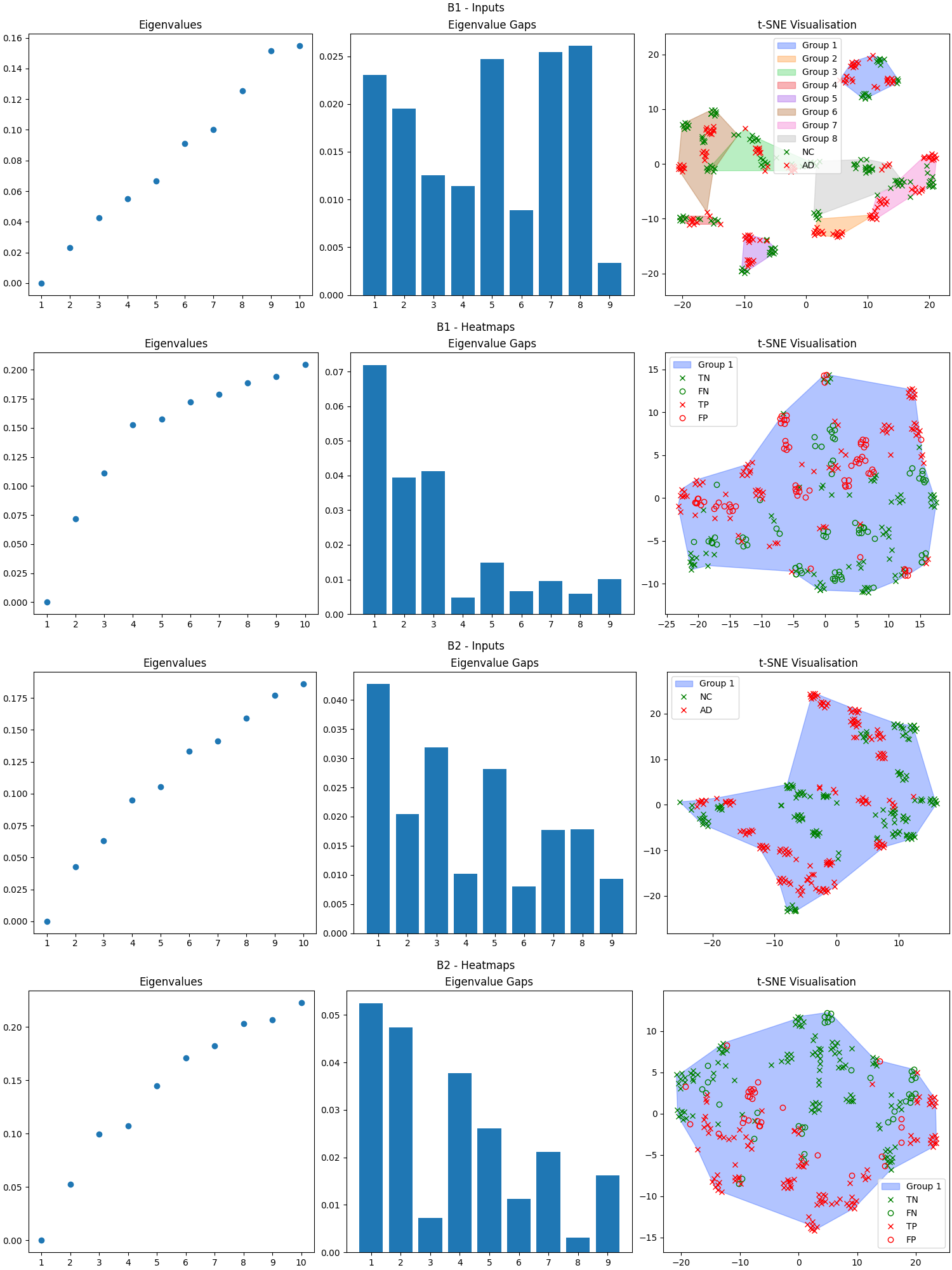}
    \caption{Eigenvalues (column 1), the respective eigenvalue gaps (column 2) of the Laplacian matrix and the t-SNE visualization (column 3) of input (rows 1 and 3) and heatmap groupings (rows 2 and 4) for models B1 (aligned T1w MRI, 13.75\% binarization) and B2 (skull-stripped T1w MRI, 13.75\% binarization).}
    \label{fig:appendix_2}
\end{figure}

\begin{figure}
    \includegraphics[width=\textwidth]{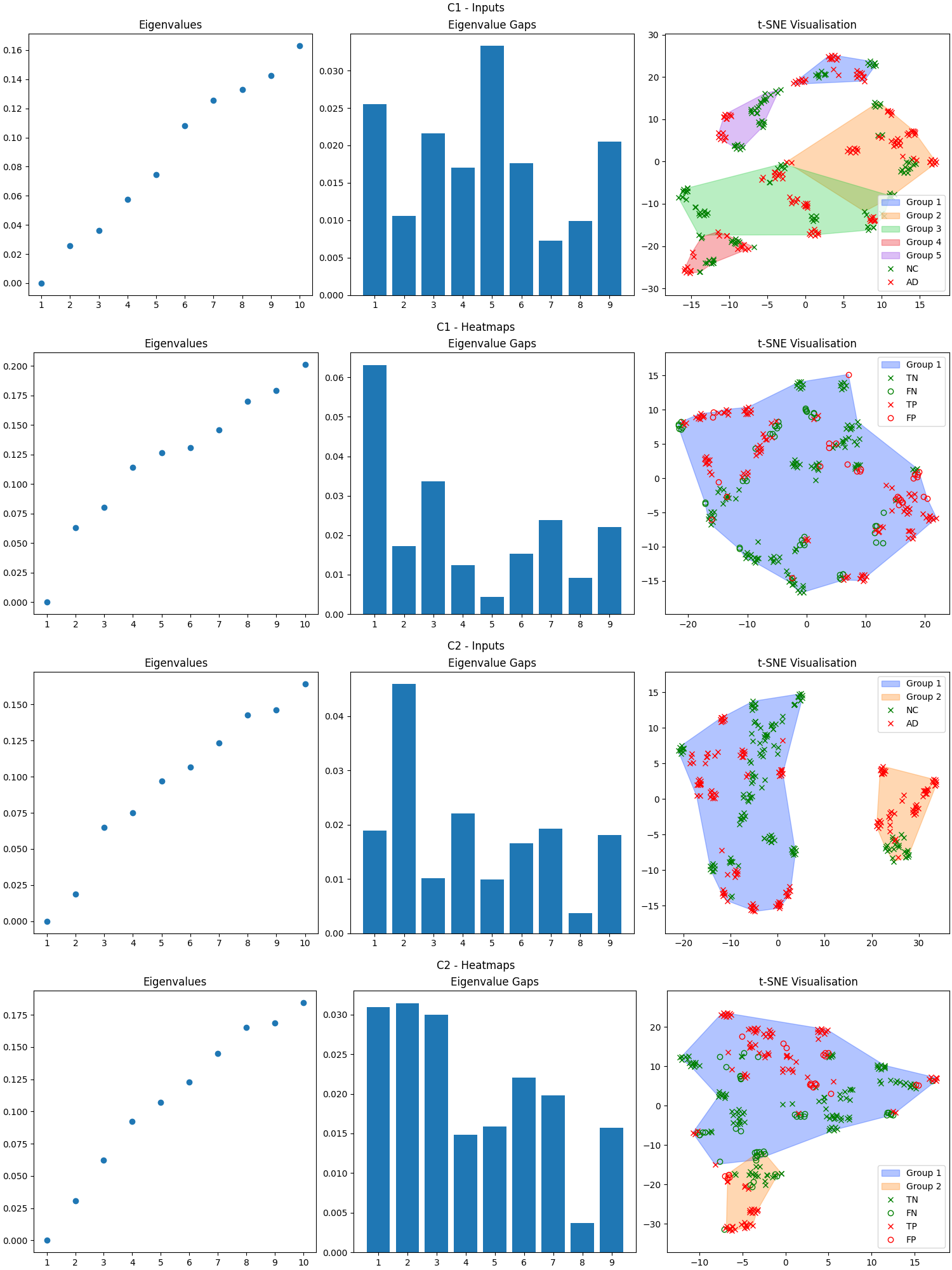}
    \caption{Eigenvalues (column 1), the respective eigenvalue gaps (column 2) of the Laplacian matrix and the t-SNE visualization (column 3) of input (rows 1 and 3) and heatmap groupings (rows 2 and 4) for models C1 (aligned T1w MRI, 27.5\% binarization) and C2 (skull-stripped T1w MRI, 27.5\% binarization).}
    \label{fig:appendix_3}
\end{figure}

\begin{figure}
    \includegraphics[width=\textwidth]{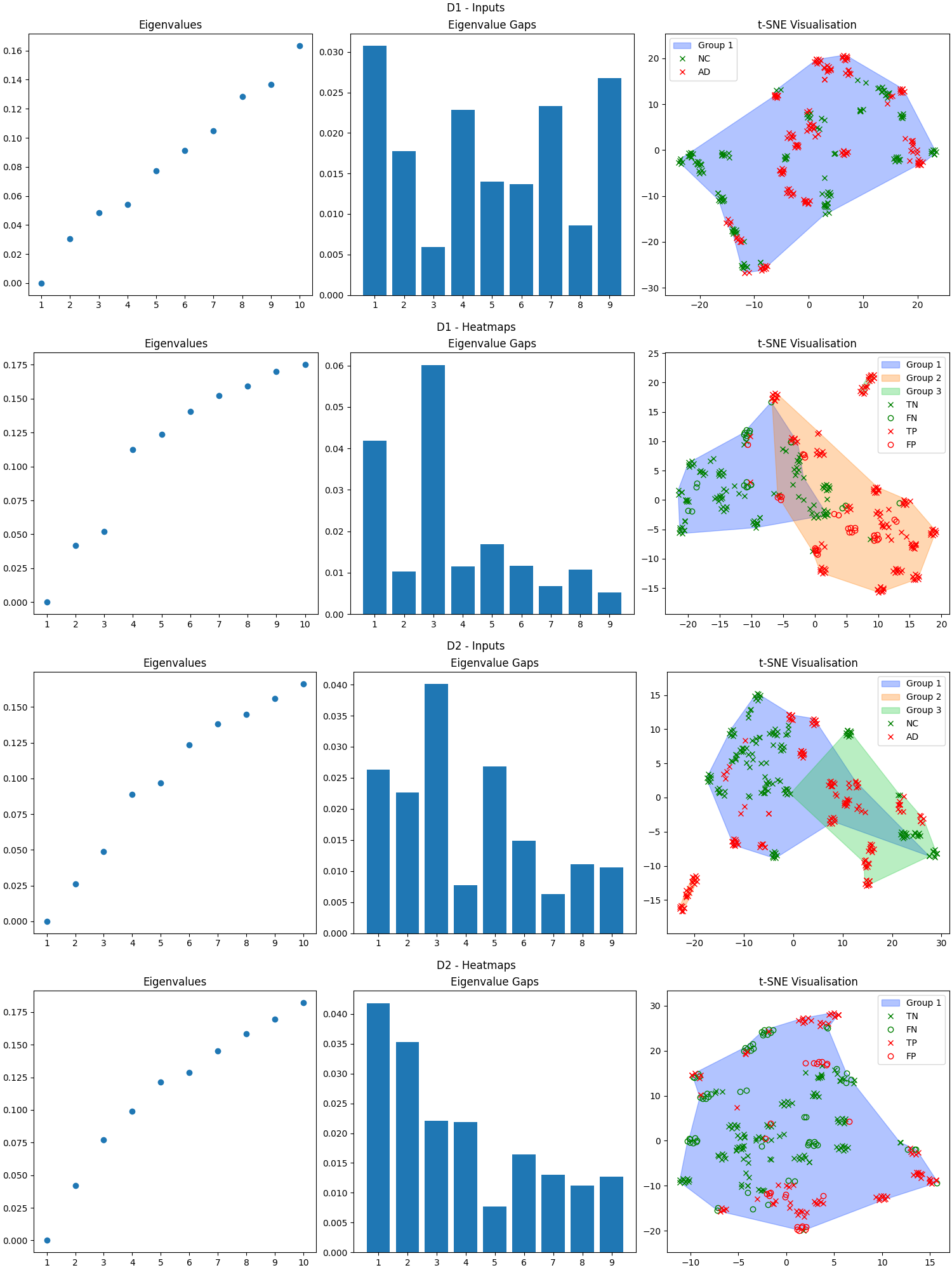}
    \caption{Eigenvalues (column 1), the respective eigenvalue gaps (column 2) of the Laplacian matrix and the t-SNE visualization (column 3) of input (rows 1 and 3) and heatmap groupings (rows 2 and 4) for models D1 (aligned T1w MRI, 41.25\% binarization) and D2 (skull-stripped T1w MRI, 41.25\% binarization).}
    \label{fig:appendix_4}
\end{figure}

\begin{figure}
    \includegraphics[width=\textwidth]{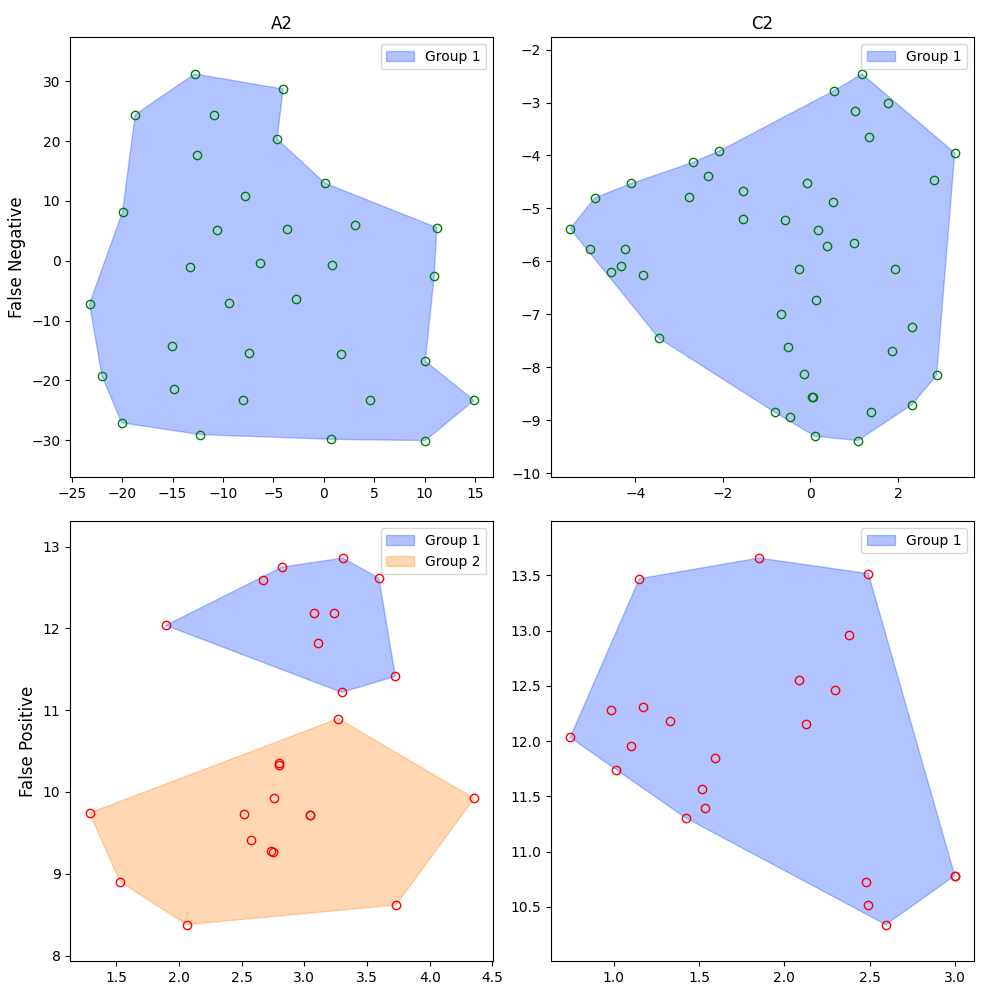}
    \caption{t-SNE visualization of heatmaps for False Negative (row 1) and False Positive (row 2) samples in models A2 (reference) and C2 (27.5\% binarization). Only the False Positive heatmaps of model A2 form two clusters, while all others group into one.}
    \label{fig:binADNI_supplemental_2_figure_1}
\end{figure}

\begin{figure}
    \includegraphics[width=\textwidth]{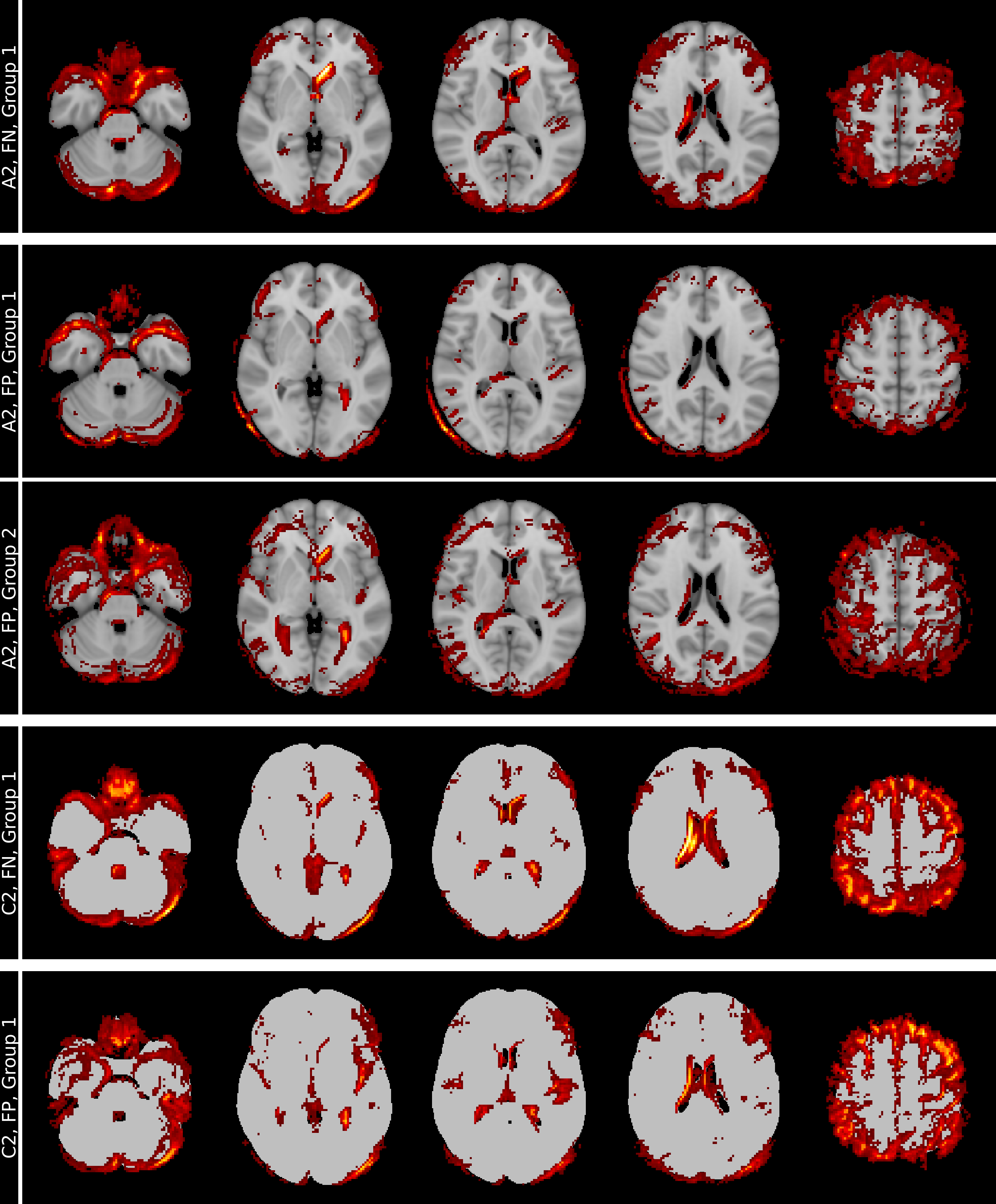}
    \caption{Mean heatmaps from spectral clustering misclassified samples in models A2 and C2. In A2 (skull-stripping), False Negative (row 1, Group 1) and False Positive (row 2, Group 1) samples show distinct relevance in the temporal lobe (column 1), while False Positive samples in row 3 (Group 2) exhibit more pronounced ventricles (column 2). Model C2 also shows differences in the temporal lobes, ventricles, and cortex between False Negative and False Positive samples (row 4 vs row 5). Note: FN = False Negative, FP = False Positive.}
    \label{fig:binADNI_supplemental_2_figure_2}
\end{figure}

The t-SNE visualization of input images and corresponding heatmaps in Figure~\ref{fig:clustering} demonstrated that only the 41.25\% binarized model without skull-stripping (D1) exhibited a heatmap grouping aligned with the underlying subject groups (NC vs. AD) while having similar classification performance as the reference model (A2). This suggests that models trained on highly preprocessed images, such as skull-stripped or lower-threshold binarized data, may introduce additional biases that alter feature utilization. The observed eigenvalue gaps (see Figures ~\ref{fig:appendix_1}, ~\ref{fig:appendix_2}, ~\ref{fig:appendix_3}, and ~\ref{fig:appendix_4}) further confirmed the presence of distinct classifier behaviors, with spectral clustering successfully distinguishing dominant and atypical relevance patterns.
\\~\\
The mean heatmaps in Figure~\ref{fig:group_heatmaps}, derived from heatmap spectral clustering groupings, highlight consistent relevance patterns, offering insight into the classifier’s decision strategies. In model D1 (41.25\% binarization without skull-stripping), the separation between AD (Group 2) and NC (Group 1) predictions is more pronounced than in models A2 (reference) and C2 (27.5\% binarization with skull-stripping). The Group 2 mean heatmap in model D1 shows distinct relevance in the left insular cortex, suggesting a more structured and positionally distinct relevance pattern, consistent with the clustering observed in the t-SNE visualization. Group 3, while similar to Group 2, also highlights regions in the skull, which is an unexpected decision strategy for classifying AD.
\\~\\
Similar patterns are found for the misclassification analysis in Figures~\ref{fig:binADNI_supplemental_2_figure_1} and ~\ref{fig:binADNI_supplemental_2_figure_2}.

\end{document}